\documentclass[%
 reprint,
superscriptaddress,
 amsmath,amssymb,
 aps,
prc,
floatfix,
]{revtex4-2}

\usepackage{graphicx}
\usepackage{aasmacros}
\usepackage{amsmath}
\usepackage{hyperref}

\begin{document}

\preprint{APS/123-QED}

\title{Total cross section of $^{14}$N+$n$ from 0.1 to 12~MeV}

\author{R.J.~deBoer}
\email{rdeboer1@nd.edu}
\affiliation{Department of Physics and Astronomy, University of Notre Dame, Notre Dame, Indiana 46556 USA}
\author{R.~Arquette}
\affiliation{Department of Physics, University of Colorado Denver, Denver, Colorado 80204 USA}
\author{D.~Bemmerer}
\affiliation{Helmholtz-Zentrum Dresden-Rossendorf, 01328 Dresden, Germany}
\author{A.~Best}
\affiliation{Università degli Studi di Napoli “Federico II”, Dipartimento di Fisica “E. Pancini”, Via Cintia, 80126 Napoli, Italy}
\affiliation{INFN - Sezione di Napoli, Via Cintia, 80126 Napoli, Italy}
\author{R.~Beyer}
\affiliation{Helmholtz-Zentrum Dresden-Rossendorf, 01328 Dresden, Germany}
\author{A.~Boeltzig}
\affiliation{Helmholtz-Zentrum Dresden-Rossendorf, 01328 Dresden, Germany}
\author{G.~Clarke}
\affiliation{Department of Physics, University of Colorado Denver, Denver, Colorado 80204 USA}
\author{J.~G\"orres}
\affiliation{Department of Physics and Astronomy, University of Notre Dame, Notre Dame, Indiana 46556 USA}
\author{T.~Hensel}
\affiliation{Helmholtz-Zentrum Dresden-Rossendorf, 01328 Dresden, Germany}
\author{A.R.~Junghans}
\email{a.junghans@hzdr.de}
\affiliation{Helmholtz-Zentrum Dresden-Rossendorf, 01328 Dresden, Germany}
\author{M.~Matney}
\affiliation{Department of Physics and Astronomy, University of Notre Dame, Notre Dame, Indiana 46556 USA}
\author{S.E.~Müller}
\affiliation{Helmholtz-Zentrum Dresden-Rossendorf, 01328 Dresden, Germany}
\author{D.~Rapagnani}
\affiliation{Università degli Studi di Napoli “Federico II”, Dipartimento di Fisica “E. Pancini”, Via Cintia, 80126 Napoli, Italy}
\affiliation{INFN - Sezione di Napoli, Via Cintia, 80126 Napoli, Italy}
\author{A.~Roberts}
\affiliation{Department of Physics, University of Colorado Denver, Denver, Colorado 80204 USA}
\author{K.~Römer}
\affiliation{Helmholtz-Zentrum Dresden-Rossendorf, 01328 Dresden, Germany}
\author{S.~Turkat}
\affiliation{Institut für Kern- und Teilchenphysik, Technische Universität Dresden, 01062 Dresden, Germany}
\author{K.~Schmidt}
\affiliation{Helmholtz-Zentrum Dresden-Rossendorf, 01328 Dresden, Germany}
\author{J.~Skowronski}
 \affiliation{Dipartimento di Fisica, Università degli Studi di Padova, 35131 Padova, Italy}
 \affiliation{INFN, Sezione di Padova, 35131 Padova, Italy}
\author{A.~Wagner}
\affiliation{Helmholtz-Zentrum Dresden-Rossendorf, 01328 Dresden, Germany}
\author{M.~Wiescher}
\affiliation{Department of Physics and Astronomy, University of Notre Dame, Notre Dame, Indiana 46556 USA}
\author{A. Yadav}
\affiliation{Helmholtz-Zentrum Dresden-Rossendorf, 01328 Dresden, Germany}
\date{\today}

\begin{abstract}
The reaction $^{14}$N$(n,p)^{14}$C is one of the main neutron poisons during $s$-process nucleosynthesis. In addition, the reaction provides insight into the yields of atmospheric nuclear weapon testing.
Because of their high level of sensitivity, total neutron cross sections provide a great deal of constraint on the modeling of reaction cross sections through the $R$-matrix analyses used for nuclear data evaluations. Yet for $^{14}$N+$n$, only one high sensitivity measurement is available and it lacks detailed information about its experimental conditions and uncertainties. With these motivations in mind, a new measurement of the $^{14}$N+$n$ total cross section has been performed at the nELBE facility. The cross sections were found to be in good agreement with previous data over much of the energy range with the key exception of the lowest energy resonance at a neutron energy of 433~keV.
\end{abstract}

\maketitle


\section{\label{sec:intro} Introduction}

The $^{14}$N$(n,p)^{14}$C reaction is a key nuclear physics ingredient in understanding neutron production for the main $s$-process (see, e.g., \citet{2016PhRvC..93d5803W}). The high cross section combined with the large amount of $^{14}$N present in the main $s$-process environment, means that it acts as an efficient neutron poison. The lowest energy strong resonance in the $^{14}$N$(n,p)^{14}$C reaction is a relatively narrow one at $E_\text{c.m.}$~=~458~keV ($J^\pi$~=~1/2$^-$, $\Gamma\approx$~8~keV). For this reason, the cross section over the low-energy region is mainly non-resonant, made up of contributions from the high-energy tails of subthreshold states and the low-energy tails of broad higher-energy resonances. At low energies (61~meV~$<E_n<$~34.6~keV), the cross section has been carefully mapped by \citet{1989PhRvC..39.1655K} who found that it was dominated by the tail contributions of subthreshold states that produce a low-energy cross section that rapidly increases towards lower energy. 
Several additional measurements have been made~\cite{1989PhRvC..39.1655K, 2016PhRvC..93d5803W, 1988ZPhyA.330..167B, 1997NIMPA.394..368S, 1995NIMPA.356..347S} that sample only a few energies but cover an energy range from 20~keV$\lessapprox E_n\lessapprox$~178~keV. The measurements over this range are in reasonably good agreement, except for those of \citet{1988ZPhyA.330..167B}, which are about a factor of two lower in cross section. There is then a gap in the experimental data between the measurements of \citet{2016PhRvC..93d5803W} and \citet{1979NSE....70..163M} (178~keV$<E_n<$~464~keV). \citet{1950PhRv...80..818J} reportedly performed measurements over this region, but the data lack uncertainties. Recently \citet{2023PhRvC.107f4617T} have remeasured over this energy range, but to compare their yield data with cross section calculations large resolution corrections need to be applied, introducing additional uncertainties. As there are many broad resonances at higher energies, whose low energy tails can conceivably contribute to the cross section over this region, interpolation over this energy region remains uncertain. 

This was recently demonstrated by \citet{2016PhRvC..93d5803W} in their comparison between their data and the JEFF-3.2 evaluation. It should be noted that the JEFF-3.2 evaluation \cite{Plompen2020} of $^{14}$N+$n$~\cite{1992ndst.book..729H} is the same as that of ENDF/B-VI.3~\cite{ENDF6_summary} and that the evaluation at low energy is based on the $R$-matrix analysis of \citet{1992ndst.book..921H} that has not been re-evaluated as of ENDF/B-VIII.0~\cite{2018NDS...148....1B}, but a revised evaluation of $n$+$^{14}$N reactions is currently underway~\cite{INDEN1, INDEN2, CM6INDEN3}.

In addition, $^{14}$N+$n$ reactions are of interest for simulating neutron transport through a variety of materials. For example, the $^{14}$N$(n,p)^{14}$C reaction is the main source of $^{14}$C in the earth's atmosphere. Neutrons for the reaction are either naturally produced from cosmic rays interacting with the atmosphere or were produced through above-ground nuclear weapons testing which significantly has influenced 
the radio-carbon method of age determination \cite{Reimer2004}. 
Using the $^{14}$N$(n,p)^{14}$C cross section and sampling the increased levels of $^{14}$C in the atmosphere, information about a nuclear explosion can be inferred (see, e.g., \citet{Burr2021}).

There have been several studies of reactions that populate the $^{15}$N system over the excitation energy range of interest. These reactions include not only $^{14}$N+$n$, but also $^{14}$C+$p$ and $^{11}$B+$\alpha$, partly because of the close proximity of their separation energies: $S_p$~=~10.207~MeV, $S_n$~=~10.833~MeV, and $Q_\alpha$~=~-10.991~MeV. For the $^{11}$B$(\alpha,n)^{14}$N reaction, there have been only a few studies~\cite{1977NuPhA.289..408N, 1975NuPhA.246...93V, 1991PhRvC..43..883W}, where that of \citet{1991PhRvC..43..883W} was the most comprehensive. Studies of the $^{11}$B$(\alpha,p)^{14}$C reaction have been made by Refs.~\cite{1976NuPhA.261..365D, 1987NuPhA.468...29T, 1991PhRvC..43..883W} and the $^{11}$B$(\alpha,\alpha)^{11}$B reaction by Refs.~\cite{1992NIMPB..64..457M, 1996NIMPB.108....1L}. Studies of $^{14}$C+$p$ reactions include measurements of $^{14}$C$(p,p)^{14}$C \cite{1968PhRv..171.1230H, 1968PhRv..172.1058H}, $^{14}$C$(p,n)^{14}$N \cite{1977NuPhA.289..408N, 1971NuPhA.173..239Y, 1951PhRv...83.1133R, 1956PhRv..104.1434S, 1959PhRv..114..571G, 1991PhRvC..43..883W}, and $^{14}$C$(p,\gamma)^{15}$N \cite{1955CaJPh..33..441B, 1971NuPhA.173..239Y, 1971NuPhA.173..239Y, 1968PhRv..172.1058H}. Finally, studies of neutron induced reactions on $^{14}$N include $^{14}$N$(n,n)^{14}$N~\cite{1955PhRv...98..728F}, $^{14}$N$(n,p)^{14}$C  \cite{1959NucPh..14..277G,1959PhRv..114..571G,1950PhRv...80..818J,1979NSE....70..163M, 2023PhRvC.107f4617T} (fast neutrons), $^{14}$N$(n,\alpha)^{11}$B \cite{1959NucPh..14..277G,1950PhRv...80..818J,1979NSE....70..163M}, and $^{14}$N$(n,\text{total})$ \cite{1952PhRv...86..483H,1951PhRv...84..775J, 1992ndst.book..729H}. Of particular note for this work is the high resolution and comprehensive measurement of the $^{14}$N$(n,\text{total})$ cross section by \citet{1992ndst.book..729H} at Oak Ridge National Laboratory.

\section{\label{sec:exp} Experimental Setup}
\begin{figure}
\centering
\includegraphics[width=8cm,clip]{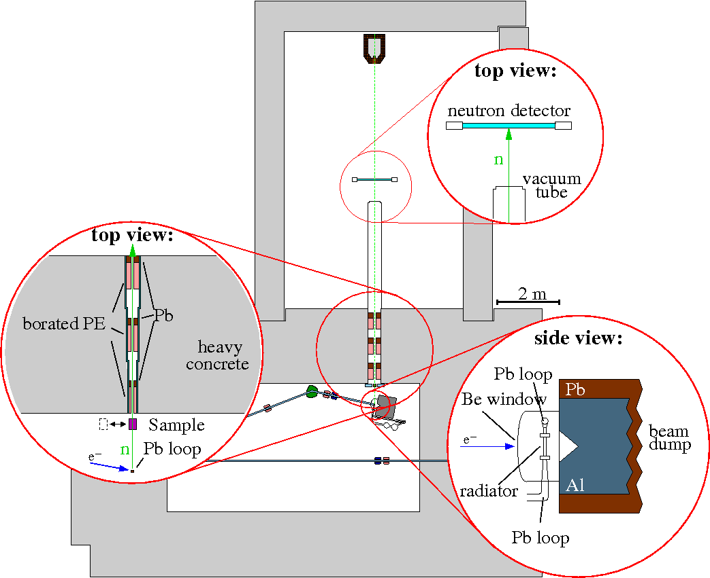}
\caption{Floor plan of the neutron time-of-flight facility nELBE at HZDR. The neutrons are produced by the electron beam hitting a liquid lead circuit as neutron producing target,
see inset on the lower right. The neutron beam is shaped by a collimator and guided to the neutron time-of-flight hall, see inset on the left. The detection setup is located in the time-of-flight hall, see upper inset. For neutron transmission experiments a plastic scintillator with a low threshold for recoil proton signals ($E_n > 10 $ keV) is used. The transmission samples are located in a movable absorber ladder in the front of the collimator.}
\label{fig:fig-1}       
\end{figure}

Helmholtz-Zentrum Dresden - Rossendorf (HZDR) operates the first photo-neutron source (nELBE) at a superconducting electron accelerator dedicated to measurements in the fast neutron range \cite{Klug2007,Beyer2013}. The floor plan of the facility is shown in Fig.~\ref{fig:fig-1}.
The superconducting electron accelerator (ELBE) accelerates electrons to a kinetic energy of typically 30~MeV in continuous-wave mode. The micro-pulse repetition rate is reduced to 101 kHz with a reduced bunch charge of approximately 10~pC to avoid an excessive bremsstrahlung intensity and neutron pulse overlap in time-of-flight measurements. The electron micro pulses have a duration of only 5-10~ps and thus allow for an excellent time resolution. A compact liquid-lead circuit is utilized as a neutron-producing target. The neutron radiator consists of a Mo-tube with a rhombic cross-section and a diameter of 11~mm through which liquid lead is pumped. The neutrons leave the neutron-producing target almost isotropically, whereas the angular distributions of electrons and bremsstrahlung are strongly forward-peaked. The collimator axis is located at an angle of 100$^\circ$ with respect to the electron beam direction. A lead absorber of 5~cm thickness mounted half-way between the neutron producing target and the collimator entrance is used to suppress the bremsstrahlung intensity (gamma flash). The target samples are mounted in a target ladder in front of the collimator entrance at a distance of 1~m from the neutron-producing target. The properties of the collimator and the neutron beam at the experimental area have been optimized in order to maintain the correlation of time-of-flight and neutron energy~\cite{Klug2007}. The collimator has a length of 2.5~m and contains three inserts of lead and borated polyethylene that are mounted inside a precision steel tube. All walls, ceiling and floor in the time-of-flight experimental hall are at least 3~m away from the neutron beam axis to help reduce the room return neutrons.   

\begin{figure}
\centering
\includegraphics[width=8cm,clip]{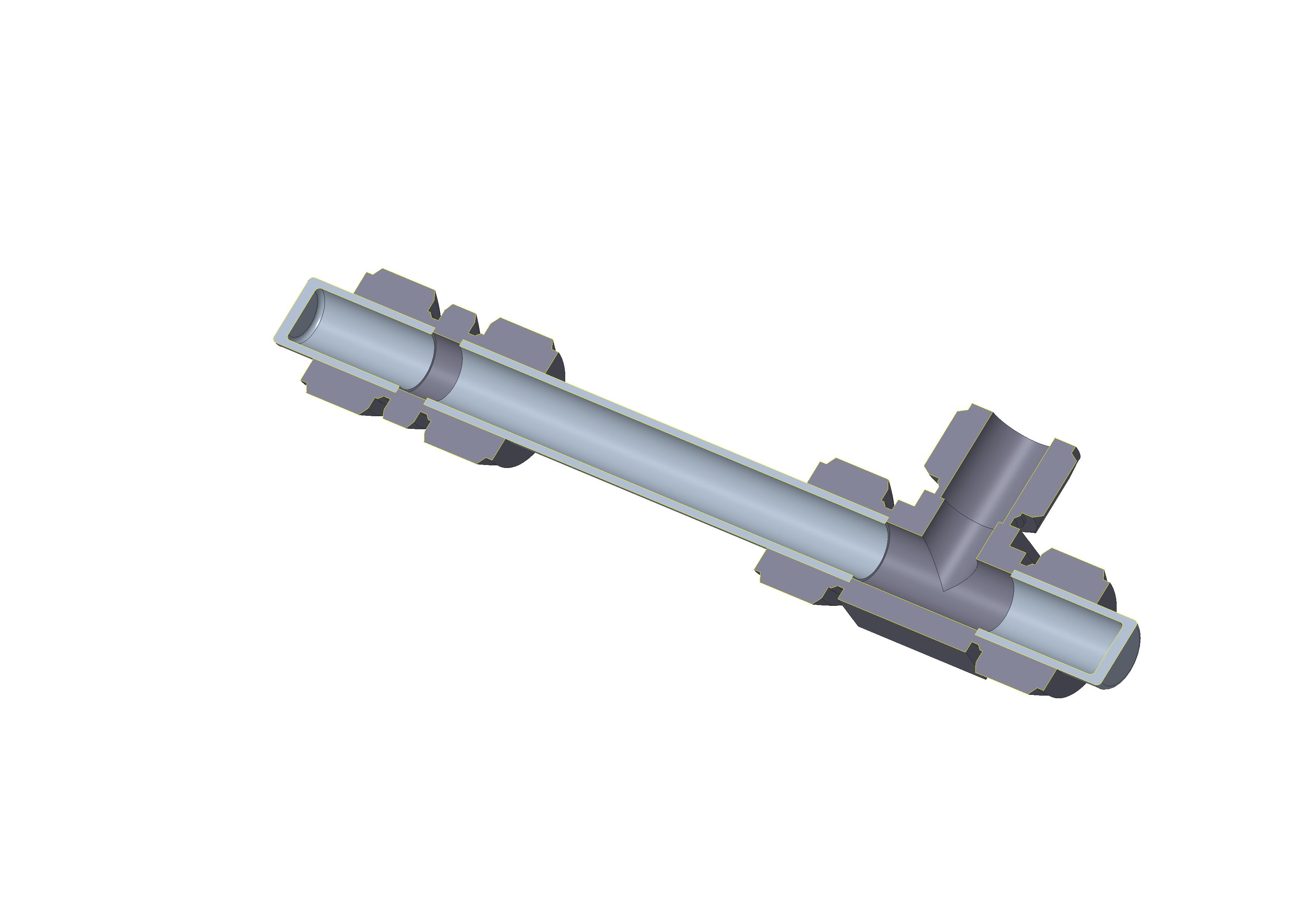}
\caption{Cross section cut of a high pressure gas cell used for nitrogen transmission measurement. The cell is made from stainless steel with cylindrical end-caps having a wall thickness of 3~mm using standard Swagelock components. The maximum pressure is up to 200~bar. The length of the gas volume is 393~mm. The tube is made from 30$\times$3~mm stainless steel. Not shown in the rendering are the valve and pressure gauge connected on the open ended flange.}
\label{fig:fig-3}       
\end{figure}

The neutrons were detected with a fast EJ-228 plastic scintillator bar with dimensions 200$\times$20$\times$5 mm$^3$ read out on both ends by high-gain PMTs Hamamatsu R2059-01. By a narrow time-coincidence window on both PMT signals, a detection threshold of about 10~keV neutron energy has been obtained~\cite{Beyer2007}. The flight path $L$ was determined to be 868.2(3)~cm long including an offset of 0.5 cm based on a comparison of measured resonance energies from transmission of N, O, Ne, Ar, and C at nELBE in comparison with the corresponding energies given in the Atlas of Neutron Resonances ref.\cite{Mughabghab2018}.  
The neutron count rate was typically 700 - 1000 per second while the bremsstrahlung pulses detected had a count rate of a 10000 - 15000 per second. The transmission measurement was made with a gaseous nitrogen sample using high-pressure gas cells made from stainless steel tubing with flat end-caps of 3~mm wall thickness. A cross section cut of the gas cell is shown in Fig.~\ref{fig:fig-3}. To cancel the transmission factor through the end caps a second identical evacuated cell was used in the ``target out of beam" measurements.

\begin{figure}
\centering
\includegraphics[width=8cm,clip]{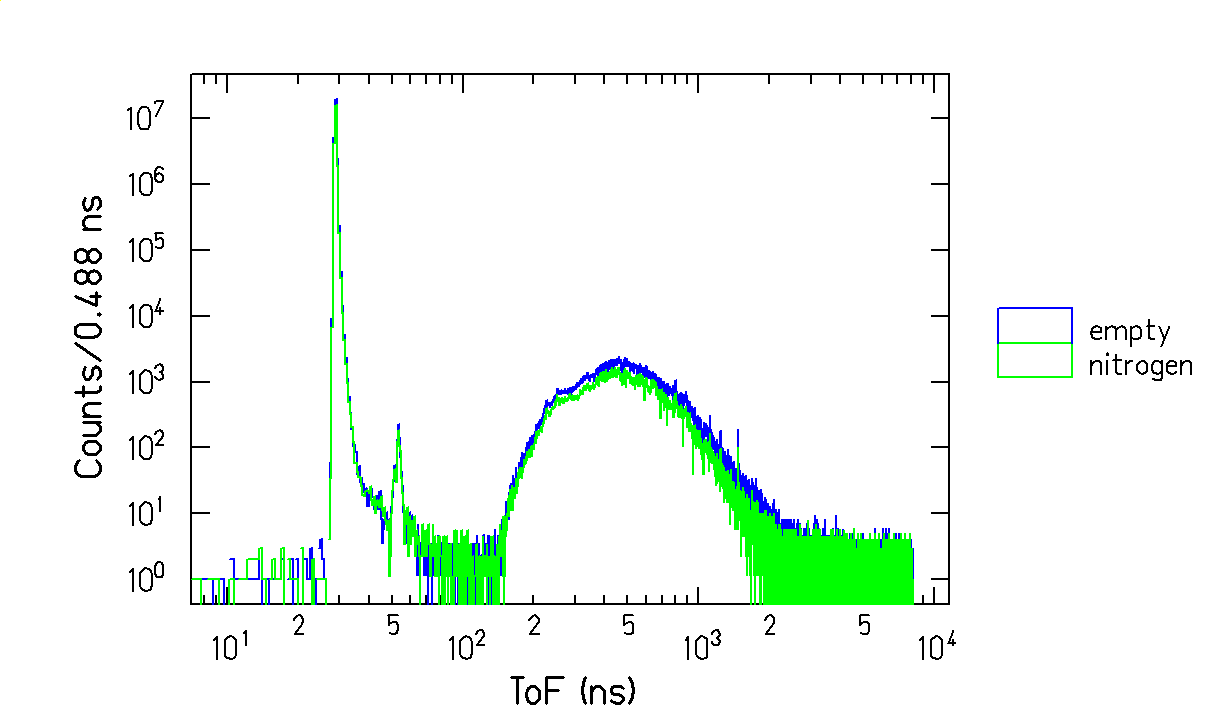} 
\caption{Time-of-flight spectra after dead-time correction with the empty gas cell and the nitrogen filled gas cell (102~bar) in the beam. The gamma-flash due to bremsstrahlung is centered at 28.945~ns with a width (FWHM) of 0.47~ns. The smaller peak to the right is due to radiation back scattered from the beam dump and rear wall. The neutron time-of-flight ranges extends from 150~ns to 2.5~$\mu$s. Random background is not subtracted in this plot.}
\label{fig:fig-3b}       
\end{figure}

The target areal density was determined by measuring the temperature and pressure during and after the filling procedure of the gas target using a high-precision pressure transducer with an absolute accuracy of 0.125~bar. The nitrogen gas with natural isotopic composition was supplied from a new high-purity bottle with a purity of 99.999\%. The fluid equation of state used to convert the measured pressure and temperature to atomic density was taken from the National Institute of Standards and Technology (NIST) data base~\cite{Linstrom2016}. The nitrogen atomic areal density of the sample was determined to be $nl=$~0.19736(24)~atoms/barn at 102.00~bar absolute pressure and a temperature of 20.4 $^\circ \rm C$. The time-of-flight of the transmitted neutrons was measured in list mode with the real-time data acquisition software package MBS (Multi-Branch-System)~\cite{MBS} developed at Gesellschaft f\"ur Schwerionenforschung (GSI), Darmstadt. The data acquisition setup consists of a single Versa Module Eurocard (VME)/ Nuclear Instrument Module (NIM) crate with a CES RIO4 front-end processor using the real-time operating system
Lynx OS. The photo-multiplier tube (PMT) signals and the accelerator frequency are fed into a
CAEN V1290A multi-hit multi-event time-to-digital converter (TDC) with (1/40.96)~ns least significant bit (LSB). This TDC is operated in trigger matching mode using a programmable
time window with 13~$\mu$s width. The time-of-flight spectrum is determined from the coincident time sum of both PMT signals relative to the accelerator frequency. A software condition is set on the time-difference of the two PMT signals to select events from the center of the beam spot on the scintillator and to reduce the detection of double hits. A dead time of 4~$\mu$s was inserted after each coincidence hit to efficiently suppress PMT afterpulses that mainly arise from the gamma-flash. The dead time was measured per event and
used to generate a time-of-flight dependent live-time factor discussed in the next section. Typical time-of-flight spectra are shown in Fig.~\ref{fig:fig-3b}.
  
\section{\label{sec:data_analysis} Data Analysis}
\begin{figure}
\centering
\includegraphics[width=8cm,clip]{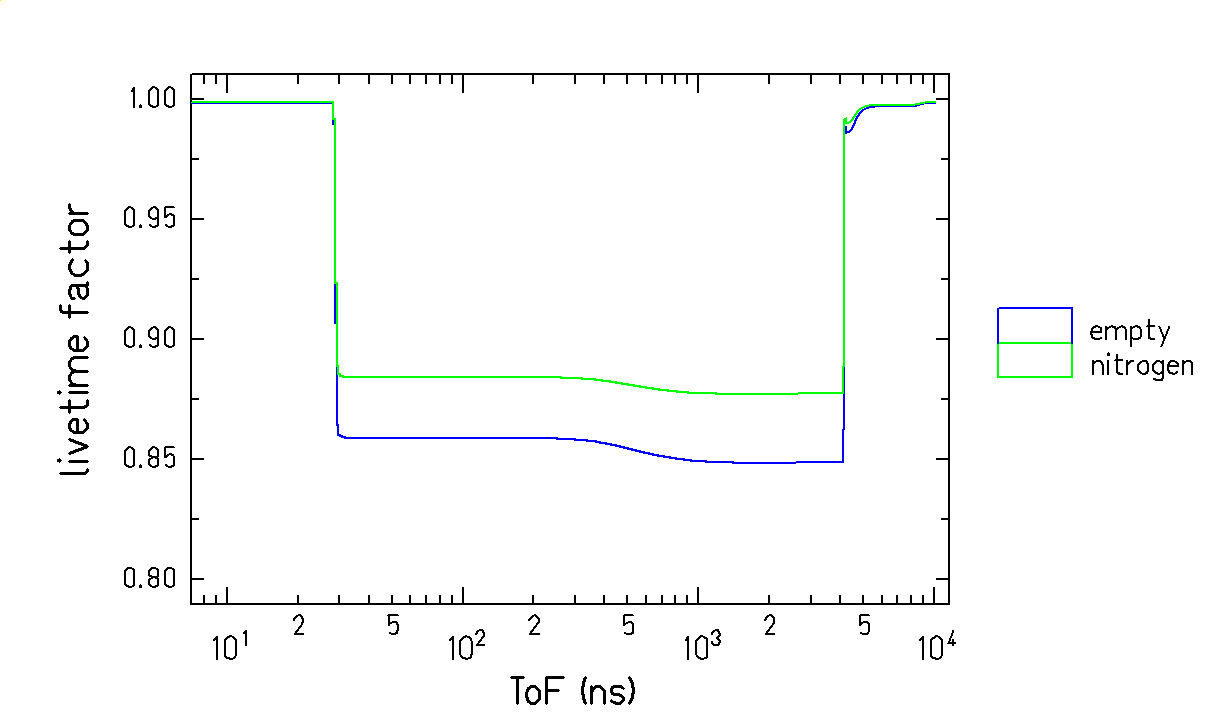}
\caption{The time-of-flight dependent live time factors for the measurements with empty gas cell and the gas cell containing nitrogen are shown on a logarithmic time scale. The nearly rectangular shape is due to the 4~$\mu$s long induced dead time after each trigger signal to suppress PMT after pulses, which are mostly due to the gamma flash signals.}
\label{fig:fig-4}      
\end{figure}

The neutron transmission $T_{exp}(t_i)$ is determined from the ratio of the background and dead time corrected count rates with nitrogen sample in the beam (in) to the empty gas cell (out) as a function of neutron time-of-flight channel $t_i$:
\begin{equation}
    \begin{split}
        T_{\rm exp}(t_i) = \frac{\sum_k(N_{{\rm in},k}(t_i) -B_{{\rm in},k}(t_i)) f_{{\rm in},k}} {\sum_k t_{{\rm real,in},k}} \\
        \quad \cdot \frac{\sum_k t_{{\rm real,out},k}} {\sum_k(N_{{\rm out},k}(t_i) -B_{{\rm out},k}(t_i)) f_{{\rm out},k}}
    \end{split}
	\label{eq:T}
\end{equation}
where $N_{\rm in/out}(t_i)$ and $B_{\rm in/out}(t_i)$ are the dead time corrected numbers of detected events and level of background events in each time-of-flight channel $i$ with and without the nitrogen target in the beam. The real times for each run $k$ with target in/out are denoted by $ t_{{\rm real,in},k}$ and $t_{{\rm real,out},k}$. The dead-time correction is time-of-flight dependent, as explained in \citet{Beyer2018}. The biggest part is caused by the gamma flash while only a smaller effect is due to the later arriving neutrons. The live time factors as a function of time-of-flight are shown in Fig.~\ref{fig:fig-4}. The absolute  time-of-flight scale was determined for each run with the measured absolute gamma-flash peak position from the TDC, the nominal flight path $L$ and the speed of light. 

\begin{figure}
\centering
\includegraphics[width=8cm,clip]{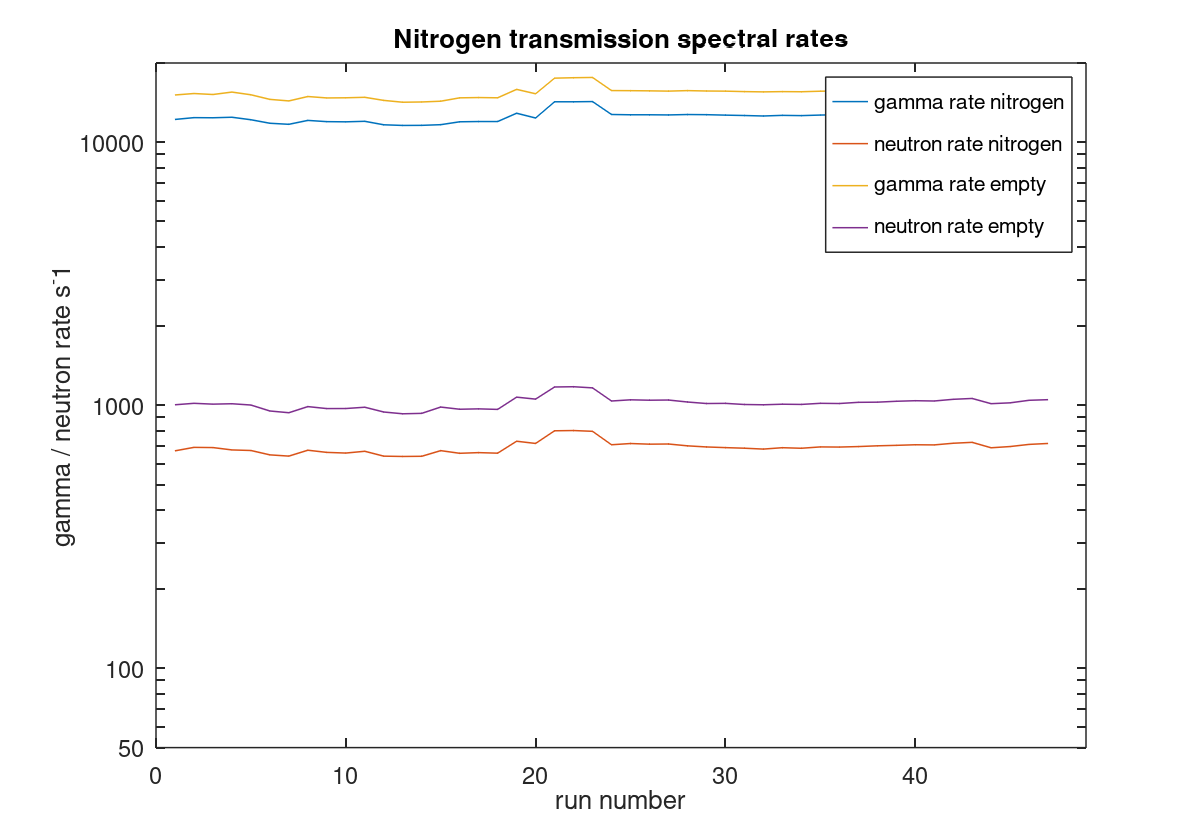}
\caption{The spectrum total count rates for the bremsstrahlung and neutron events are shown for subsets of the data (runs) lasting 2-4 hours typically including several target-in and out sample changes.}
\label{fig:fig-8}       
\end{figure}

The transmission experiment was done by periodically moving the nitrogen sample and empty gas cell in and out of the beam with counting times adapted to maximize the statistics. Typical counting times for each setting were 10~min empty and 20~min with nitrogen for a total beam time of 156~hours. Listmode data were recorded in separate runs of about 2-4~hours. The spectrum average neutron and gamma-count rates for each run are plotted in Fig.~\ref{fig:fig-8} to illustrate the average beam stability in the measurement. The relative beam stability
(standard deviation / mean value) of the average count rates of the separate runs amounts to 0.05. 

\begin{figure}
\centering
\includegraphics[width=8cm,clip]{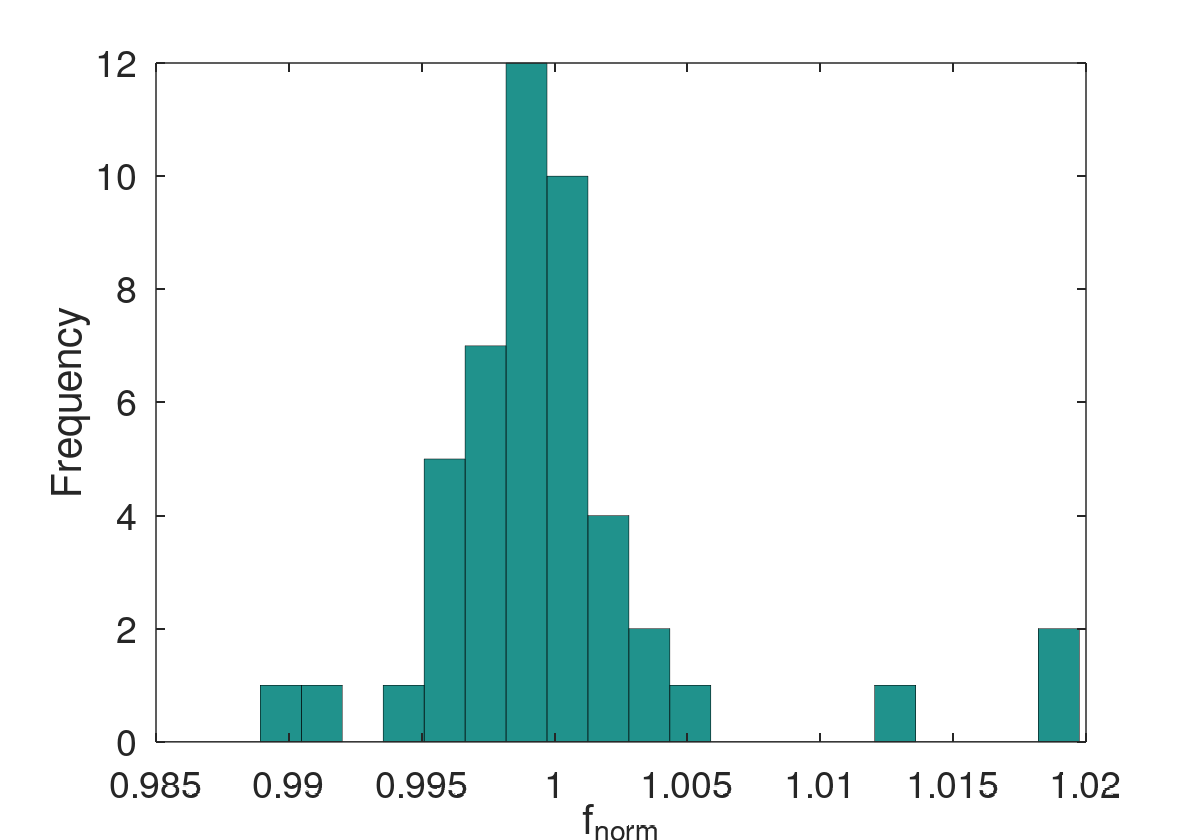}
\caption{The distribution of $f_{in,k}/f_{out,k}$, see Eqs.~(\ref{eq:f_out}) and (\ref{eq:f_in}), is shown for subsets of the data (runs) lasting 2-4 hours typically including several target-in and out sample changes.}
\label{fig:fig-9}       
\end{figure}

The transmission was calculated using the sum of all these runs $k$ where no beam intensity fluctuation or failure occurred in the ``sample in" and ``empty gas cell" settings. The factors $f_{{\rm in},k}$ and $f_{{\rm out},k}$ are normalization factors to correct for the remaining fluctuations in the neutron source intensity:
\begin{equation}
f_{{\rm out},k} =  \frac{t_{{\rm real,out},k}}{\sum_i (N_{{\rm out},k}(t_i) - B_{{\rm out},k }(t_i))}
\label{eq:f_out}
\end{equation}
\begin{equation}
f_{{\rm in},k} = \langle T \rangle  \frac{t_{{\rm real,in},k}}{\sum_i (N_{{\rm in},k}(t_i) - B_{{\rm in},k }(t_i))}
\label{eq:f_in}
\end{equation}
The neutron count rate  for each run $k$ with target out of the beam summed over the full neutron time-of-flight range $i$ is used as a neutron source intensity monitor in Eq.~(\ref{eq:f_out}). For the ``target in" settings, the neutron source intensity monitor is approximated with the target-in count rate divided by the mean integral transmission factor $\langle T \rangle$ over the full neutron time-of-flight range $i$ of the full experiment including all runs $k$. The distribution of the normalization factors $f_{in,k}/f_{out,k}$, see Eqs.~(\ref{eq:f_out}) and (\ref{eq:f_in}), is shown in Fig.\ref{fig:fig-9}. The mean value is one by definition and the standard deviation is 0.005.

From the measured transmission as a function of time-of-flight $T_{\rm exp}(E_{\rm n}(t_i))$, an effective neutron total cross section $\langle\sigma_{\rm tot}(E_{\rm n})\rangle$ can be determined:
\begin{equation}
  \langle\sigma_{\rm tot}(E_{\rm n})\rangle = -\frac{1}{nl} \ln{T_{\rm exp}}
  \label{eq:sigma}
\end{equation}
where $nl$ is the atomic areal density of the target sample.

\begin{figure}
    \centering
    \includegraphics[width=0.5\textwidth]{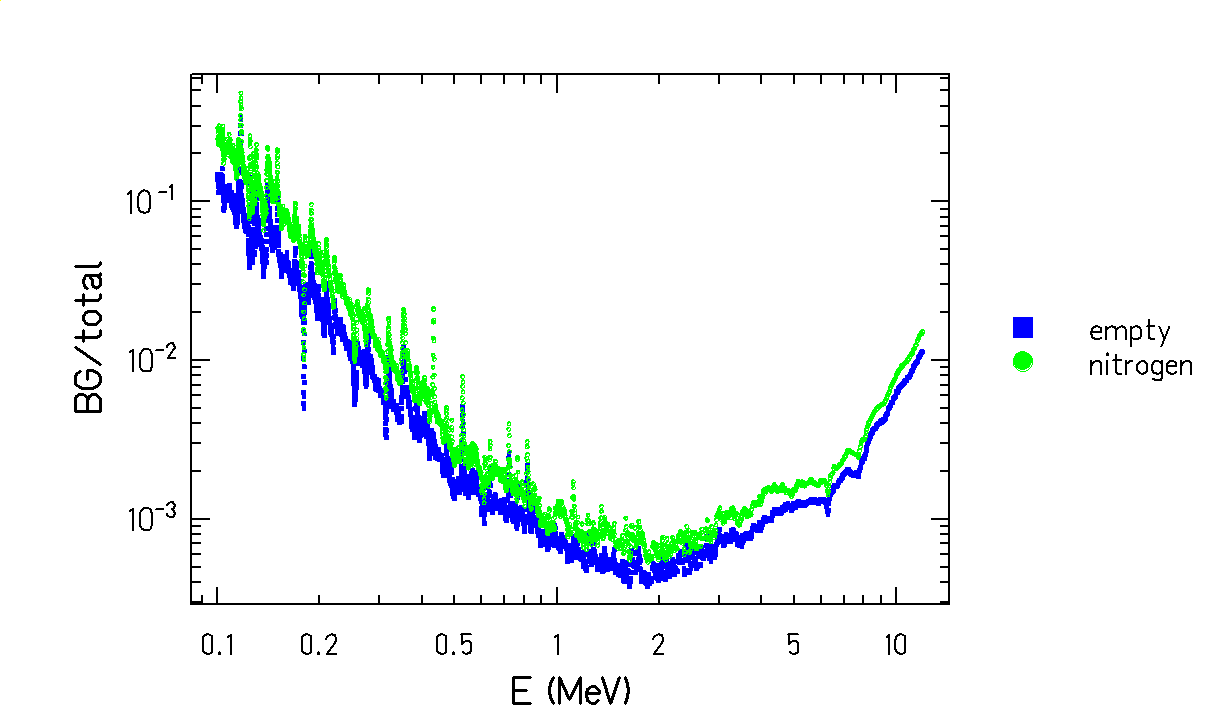}
    \caption{The background to total counts ratios are shown for the full data set of the transmission measurement. The green data points (circles) were measured with the nitrogen sample in the beam, the blue data points (squares) with the empty gas cell. The
    statistical uncertainties have not been plotted to improve the visibility of the structures in the spectra. }
    \label{fig:background-total}
\end{figure}

To determine the neutron transmission and the total cross section from the measured time-of-flight distribution with a relative accuracy of a few percent, several corrections need to be considered:
\begin{enumerate}
\item Correction for a time-of-flight dependent dead time,
\item Subtraction of a random background in the time-of-flight spectra,
\item Correction for fluctuations of the neutron-beam intensity,
\item Correction for in-scattering of neutrons,
\item Correction for resonant self-shielding in thick transmission samples.
\end{enumerate}
Random background and dead-time corrections are important at low and high neutron energy, where the neutron source intensity is already decreasing. In this experiment a low beam intensity and a very compact neutron-producing target without any materials that would slow down neutrons were used. There is no neutron moderator in the target setup and the amount of in-beam gamma-rays from neutron capture is negligible. The random background can be described by a constant value in time-of-flight spectrum. It is mostly dominated by random coincidences
due to ambient natural radioactivity, whereas the room return background of neutrons is relatively low~\cite{Hannaske2013}. The determination of the background levels $B_{\rm in/out}$ is done by calculating
the mean bin content in the time-of-flight ranges before the $\gamma$-flash and between 4.68 to 7~$\mu$s. In the time-of-flight region around 150~ns, before the first fast neutrons arrive, the background level is higher than estimated from the mean bin content between 4.68 to 7~$\mu$s. This background tail is due to insufficient suppression of PMT after-pulses mostly from the gamma-flash. Due to the low background-to-total ratio, it has been neglected in the transmission determination~\cite{Beyer2018}. The background-to-total ratio is shown in Fig.~\ref{fig:background-total}. As the background is taken as a constant value from outside of the neutron time-of-flight range, the spectra show peaks and dips which are related to the pure source spectrum, which contains known peaks from near-threshold photo neutron production in lead and from the resonant structure of the nitrogen and other layers of matter in the beam. In the neutron energy range below 200 keV the neutron intensity from the source decreases and uncertainties in the background shape get large enough to influence the transmission measurement. 
The uncertainties of the transmission measurement and total cross-section determination have already been discussed in \citet{Hannaske2013} and \citet{Beyer2018}. The important uncertainties will be briefly revisited here. The fluctuations of the neutron-beam intensity were measured during the sample cycling and found to have a small influence. From transmission values determined from subsets of the data, a fluctuation of less than 0.5\% was found and included in the total systematic uncertainty. In-scattering of neutrons was minimized by the geometry of the setup: The collimator strongly limits the solid angle under which neutrons can be registered and only neutrons passing through the full length of the sample can hit the detector. The probability for multiple in-scattering of fast neutrons has been found to be less than 0.1\%~\cite{Hannaske2013}.

\begin{figure}
    \centering
    \includegraphics[width=0.5\textwidth]{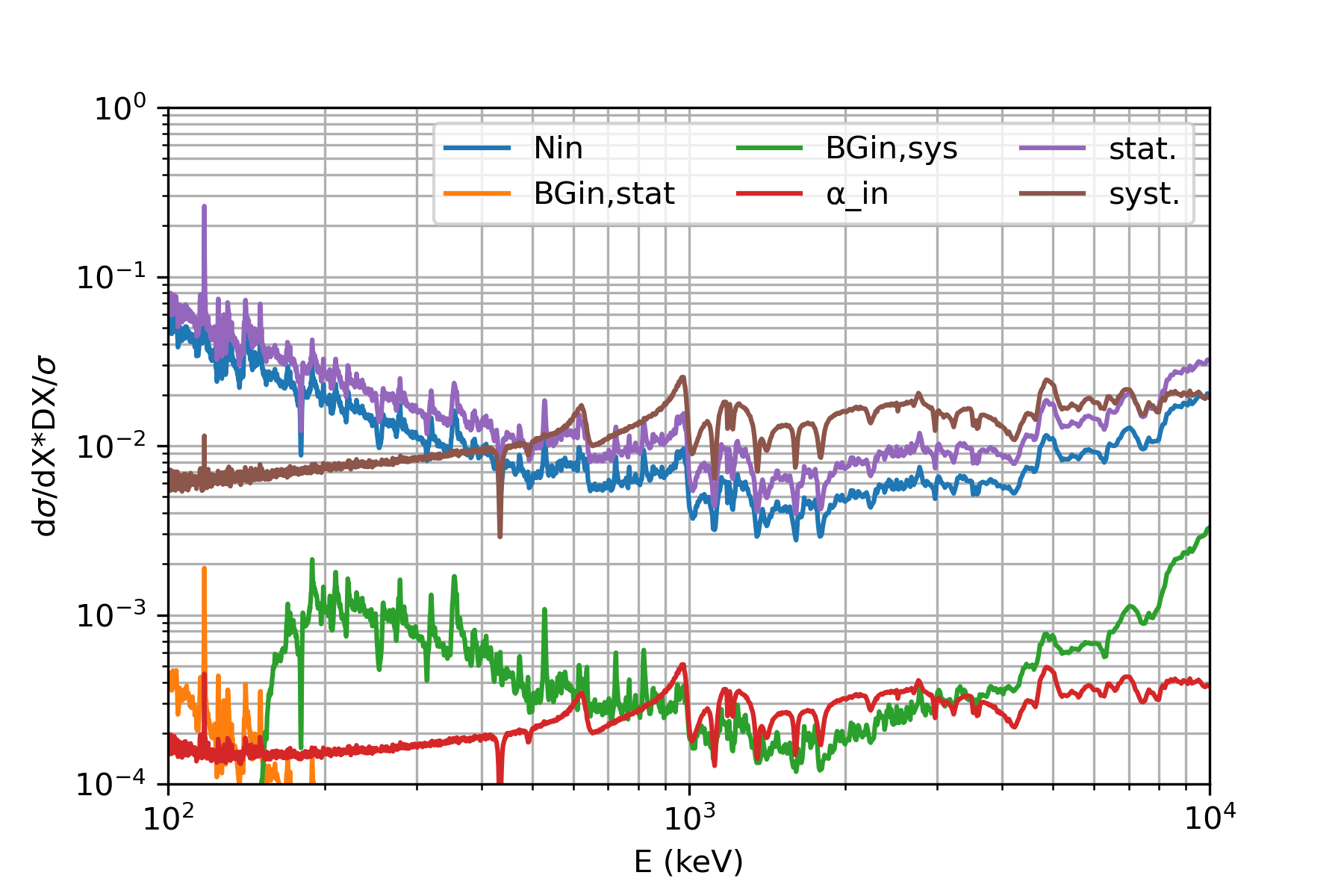}
    \caption{The uncertainty budget of the measured total cross section of nitrogen. The relative contribution of each parameter $X \in \{  N_{in}, \alpha_{in}, BG_{in} \} $ to the uncertainty of the total cross section $\sigma$ due to statistical (stat.) and
    systematic (sys.) effects is shown for a time binning of 1~ns. The total statistical (purple) and systematic (brown) uncertainty also contain the contributions from the target out data.}
    \label{fig:fig-5}
\end{figure}

In Fig.~\ref{fig:fig-5} the uncertainty budget of the total cross section is shown for the nitrogen sample measurements. The plot looks similar for the empty target measurements. To quantify the contribution of each parameter $X \in \{ N_{\rm in}, \alpha_{\rm in}, BG_{\rm in} \}$ used to determine the total cross section $\sigma$ the quantity $\frac{\partial \sigma}{\partial X} \frac{\Delta X}{\sigma} $ was calculated by differentiating Eq.~(\ref{eq:T}). These quantities are plotted for a time-of-flight bin size of 1~ns. It is visible that the uncertainty of the background level due to counting statistics ($BG_{\rm in/out,stat}$) is negligible for the overall uncertainty due to statistical effects $\frac{\Delta\sigma}{\sigma}$(stat.). The contribution of the uncertainty of the background level due to the inclusion of the energy-dependent uncertainty ($BG_{\rm in/out,sys}$) to the overall uncertainty due
to systematic effects $\frac{\Delta\sigma}{\sigma}$(sys.) becomes noticeable at the upper and lower ends of the neutron energy spectrum available at nELBE. Nevertheless, over the complete energy range, the uncertainty of the flux normalization factor $f_{norm}$ is dominating the systematic uncertainty. The uncertainties of the samples areal density ($nl$) and the dead time correction factor ($\alpha$) are less important. In total, $\frac{\Delta\sigma}{\sigma}$(sys.) sums up to about 1.5\%.

\begin{figure*}
    \centering
    \includegraphics[width=1.0\textwidth]{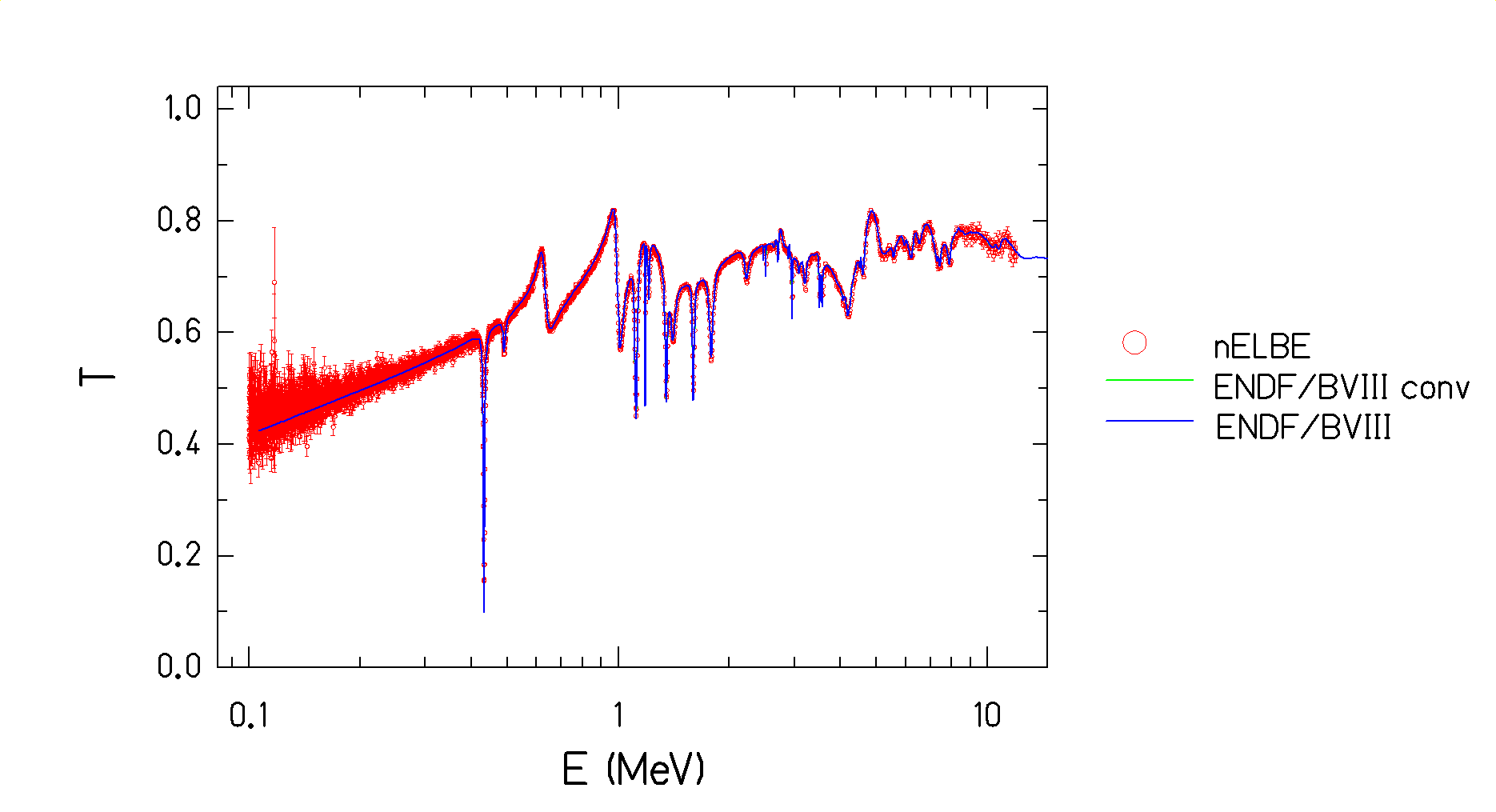}
    \caption{The measured neutron transmission of nitrogen
    with an areal density of 0.19736(24)~atoms/barn in the full energy range of 0.1 to 12~MeV. The time-of-flight bin size is 0.488~ns. The experimental data are compared with calculated transmission from ENDF-B/VIII.0 point wise total cross sections 
    (blue) and a transmission curve including the experimental resolution function (green).  
    }
    \label{fig:fig-6}
\end{figure*}

\begin{figure*}
    \centering
    \includegraphics[width=1.0\textwidth]{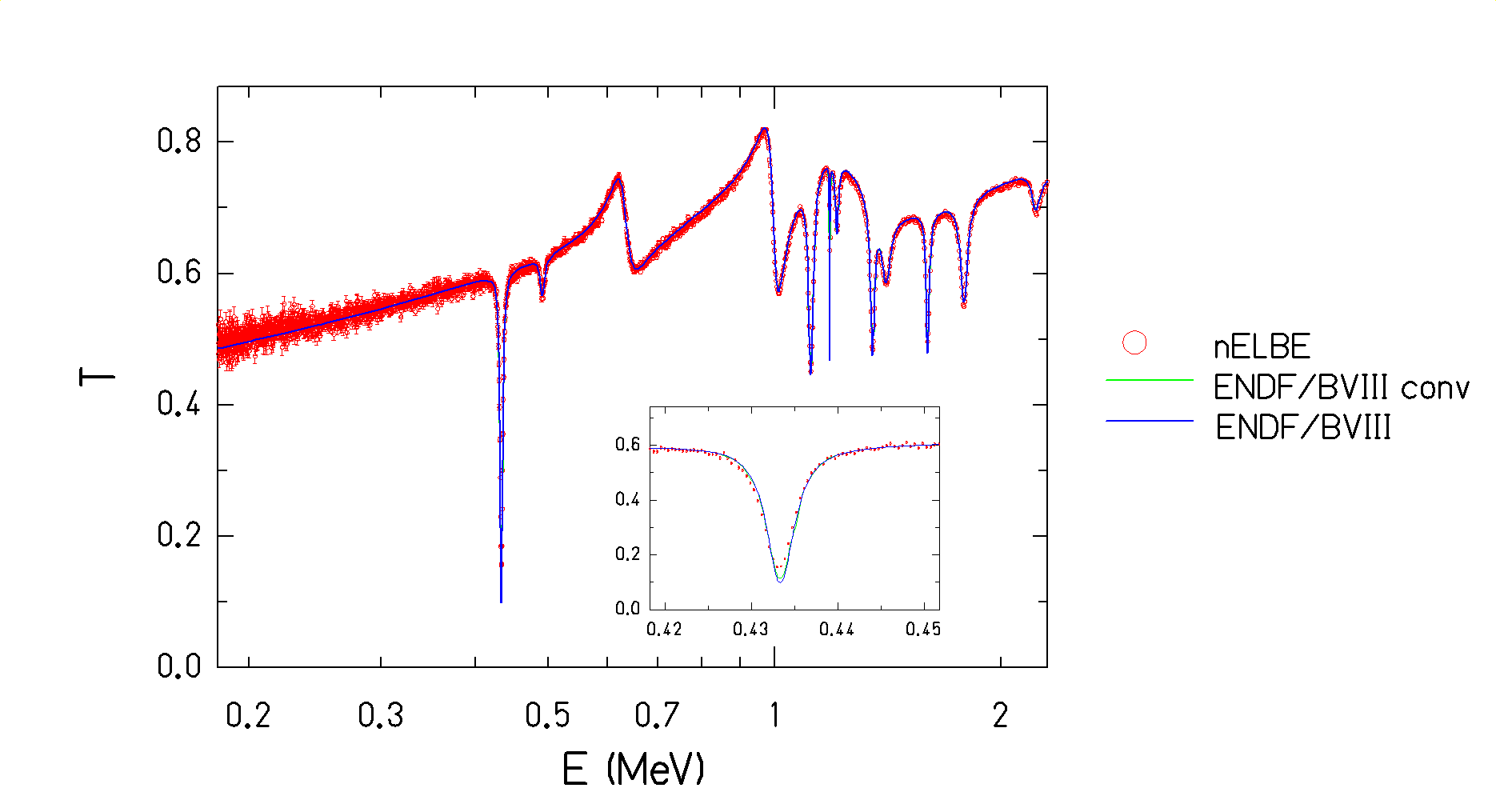}
    \caption{The measured neutron transmission of nitrogen in the energy range below
    2~MeV, description of data as in Fig.~\ref{fig:fig-6}. The experimental resolution
    is sufficient to resolve most known resonances with the exception of the resonance 
    at 1184~keV, which has a total width of 1.4~keV.
    The inset shows the resonance at 433 keV. The calculated transmission based on ENDF/B VIII is too small at the resonance peak.}
    \label{fig:fig-7}
\end{figure*}

The measured neutron transmission of nitrogen is shown in Fig.~\ref{fig:fig-6} and Fig.~\ref{fig:fig-7}  compared to a calculated transmission curve based on the evaluated cross section from ENSDF/B~VIII.0. The measured transmission extends through the fast neutron energy range from about 0.1 to 12~MeV. The nitrogen thickness was chosen in order not to saturate the first resonance at 433~keV neutron kinetic energy. The energy resolution of the time-of-flight measurement
determines how well the resonant structure of the underlying total cross section can be  observed. 

The resolution function of nELBE is modeled as a Gaussian function 
\begin{equation}
    R(E-E') = \frac{1}{\Delta E\sqrt{2\pi}} \exp\left(-\frac{1}{2}\frac{(E-E')^2}{(\Delta E)^2}\right)
\end{equation}
with the RMS width $\Delta E$, which consists of the time resolution of the plastic scintillator $\Delta t$ used for detecting the transmitted neutrons and the standard deviation of the nominal flight length $\Delta L$ due to the combined thickness of the neutron producing target and the plastic scintillator $\Delta L$ 
\begin{equation}
    \frac{\Delta E}{E} =  (\gamma+1) \frac{\Delta v}{v}
\end{equation}
\begin{equation}
    \frac{\Delta v}{v} = \sqrt{\left(\frac{\Delta t}{t}\right)^2 + \left(\frac{\Delta L}{L}\right)^2}
\end{equation}
where $\gamma$ is the Lorentz parameter for a neutron with kinetic energy $E$ and velocity $v$. The time resolution of the plastic scintillator was determined experimentally from the measured width of the bremsstrahlung peak. As the electron beam pulse length is only 5-10~ps, the measured width corresponds closely to the time resolution ($\Delta t $= 0.20~ns, 1~std. dev.). The width of the neutron producing target (11~mm) and the plastic scintillator (5~mm) give $\Delta L = \sqrt{((1.1 \mbox{ cm})^2 + (0.5 \mbox{ cm})^2)/12}$=~0.35~cm, 1 std. dev, where division by $\sqrt{12}$ converts the FWHM of the rectangular flight path distribution to the standard deviation of an equivalent Gaussian. The transmission curve including the experimental resolution with the ENDF/B-VIII.0 evaluated cross section has been calculated by numerical integration of 
\begin{equation}
    T(E) = \int R(E-E') \exp(-nl \sigma_{ENDF}(E'))dE'.
\end{equation}

\begin{figure}
    \centering
    \includegraphics[width=0.5\textwidth]{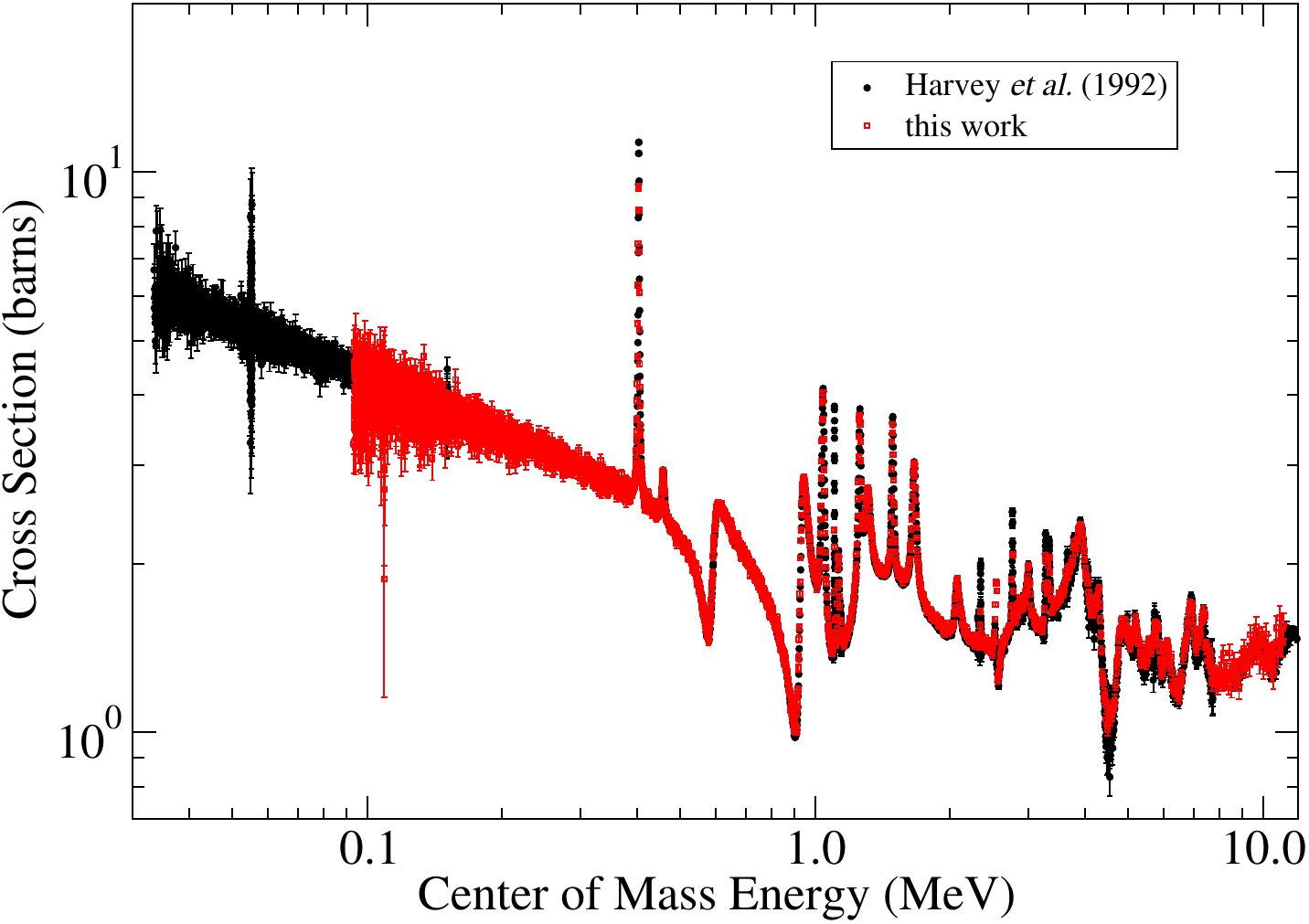}
    \caption{Comparison of the $^{14}$N total cross section data of this work with that of \citet{1992ndst.book..729H}.The data sets are largely consistent when experimental resolution is applied.}
    \label{fig:compare_full_energy_range}
\end{figure}

\section{\label{sec:r-matrix} $R$-matrix Analysis}

$R$-matrix calculations were performed using the code \texttt{AZURE2}~\cite{2010PhRvC..81d5805A}, representing a continued analysis of the $^{15}$N system as presented previously in \citet{2020JPhCS1668a2011D} and \citet{PhysRevC.108.035809}. The $R$-matrix calculations utilize the alternative parameterization of \citet{2002PhRvC..66d4611B} so that observable resonance parameters can be used directly as inputs for the $R$-matrix calculations instead of formal ones~\cite{1958RvMP...30..257L}. It also eliminates the need for boundary condition parameters, leaving the channel radius as the sole model parameter. The typical simplification has been made that the same channel radius is used for all channels of a given particle partition. The $\alpha$-particle, proton and neutron channel radii were 5.5, 5.0, and 4.0~fm, respectively.

\begin{figure*}
    \centering
    \includegraphics[width=1.0\linewidth]{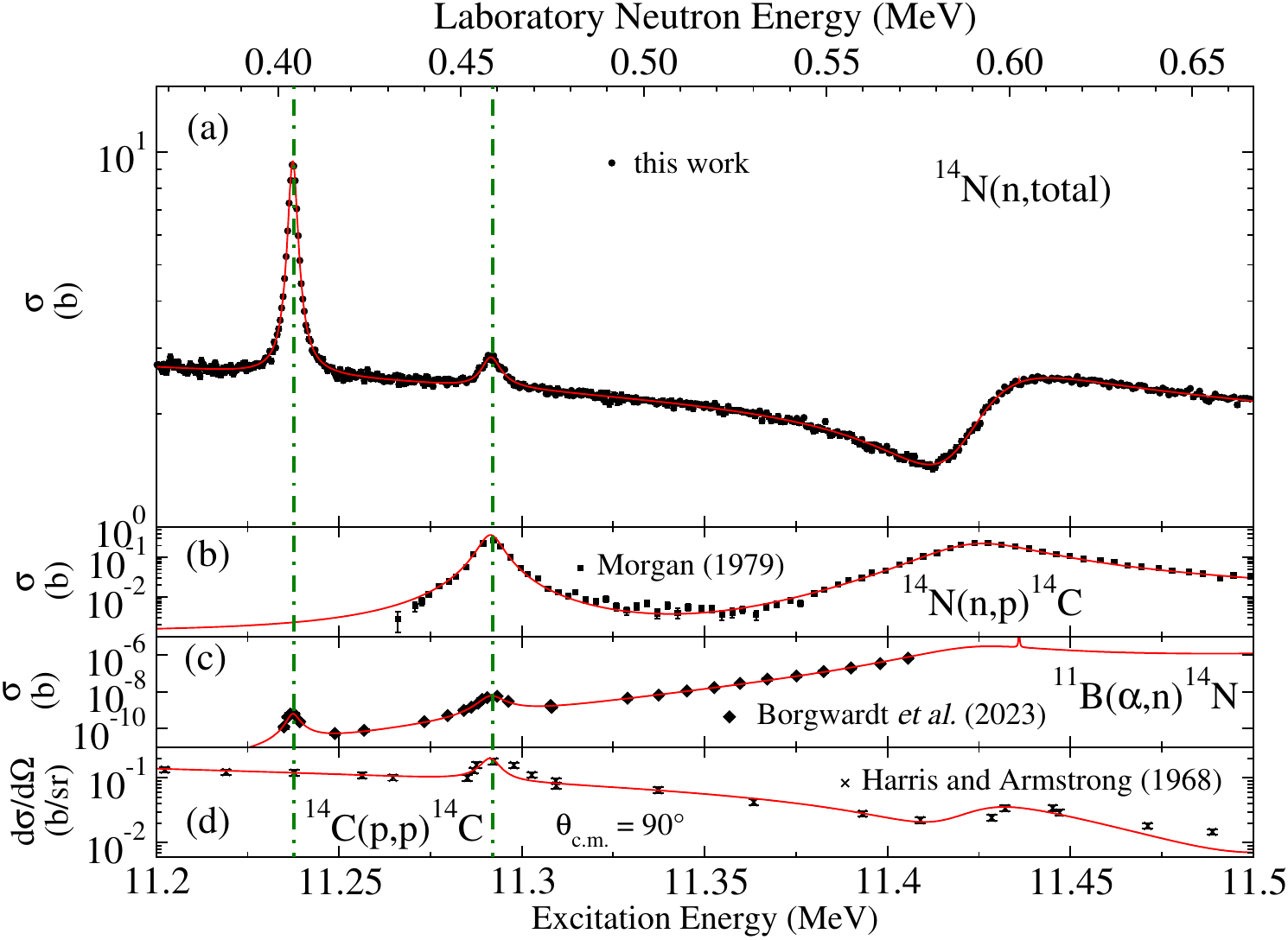}
    \caption{Multichannel $R$-matrix fit (solid red line) that includes (a) the present total neutron cross section data, (b) the $^{14}$N$(n,p)^{14}$C data of \citet{1979NSE....70..163M}, (c) the $^{11}$B$(\alpha,n)^{14}$N data of \citet{PhysRevC.108.035809} and (d) the $^{14}$C$(p,p)^{14}$C data of \citet{1968PhRv..171.1230H}.}
    \label{fig:multichannel_fit}
\end{figure*}

Although the present total neutron cross section measurements extend up to $E_n\approx$~12.0~MeV, the analysis concentrates on the energy region below $E_n <$~1~MeV in order to both simplify the $R$-matrix analysis and because the focus is to investigate the properties of the $E_n$~=~433~keV resonance observed in the total neutron cross section data. Over this energy range, there are only three particle partitions that are energetically accessible: $n$+$^{14}$N, $p$+$^{14}$C, and $\alpha$+$^{11}$B, all only accessing the ground state. Since, for several resonances, the corresponding compound nucleus levels have proton and $\alpha$-particle partial widths with similar strength compared to the neutron partial width, the inclusion of these other partitions are needed to reproduce the total neutron cross section. To constrain the branchings of these partial widths in the fit, representative data sets~\cite{PhysRevC.108.035809, 1989PhRvC..39.1655K, 1968PhRv..171.1230H, 1979NSE....70..163M} from these other reaction channels are also included to improve the fidelity of the fit. The fit is shown in Fig.~\ref{fig:multichannel_fit}.

To estimate uncertainties for the $R$-matrix fit, the Bayesian $R$-matrix Inference Code Kit (\texttt{BRICK})~\cite{2022FrP....10.8476O} was utilized. \texttt{BRICK} is a Python package that serves as a mediator between the \texttt{AZURE2} $R$-matrix code and the MCMC sampling software \texttt{emcee}~\cite{2013PASP..125..306F}. Uncertainties for the level parameters as well as those for the cross sections that were fit were extracted. The Bayesian uncertainty analysis included the systematic uncertainties of each data set as a Gaussian prior. More information can be found in the Supplemental Material~\cite{supplemental_material}.

\section{\label{sec:dis} Discussion}

When the resolution function for the present measurements is applied to the $R$-matrix cross section, there is good agreement between the data of this work and those of \citet{1992ndst.book..729H} over nearly the entire overlapping energy range.  
The comparison of the experimental total cross section measured in this work with \citet{1992ndst.book..729H}  in Fig.~\ref{fig:compare_full_energy_range} shows already this good agreement. 
The exceptions are in the shapes of some of the narrow resonances (1184 keV), where the most discernible differences are in the first resonance at $E_n$~=~433~keV. This resonance was first observed in total cross section measurements in 1951 by \citet{1951PhRv...84..775J}, but its level properties proved difficult to constrain due to its narrow width. In both the early measurements of \citet{1951PhRv...84..775J} ($E_n$~=~433~keV) and \citet{1952PhRv...86..483H} ($E_n$~=~430~keV, $\Gamma$~=~3.5~keV), the total width was reported to be similar to the experimental resolution of a few keV and that $J>1/2$. After these two early total cross section measurements, no new measurements were made until those of \citet{1992ndst.book..729H} in 1992. That measurement obtained significantly improved energy resolution and uncertainty giving $E_n$~=~433.35(3)~keV, $\Gamma_n$~=~2.46(4)~keV, $\Gamma_\gamma$~=~40(70)~eV and $J=7/2$. In particular, their significantly improved energy resolution resulted in the first firm spin assignment for this resonance, as the resonance height is indicative of the spin of the corresponding level.

Although it is a small effect, the peak cross section produced by a 7/2$^+$ resonance in the $R$-matrix fit slightly overshoots the two highest data points of \citet{1992ndst.book..729H}. To resolve this disagreement, \citet{1992ndst.book..729H} included a $\gamma$-ray partial width of 40(70)~eV to damp the height of the resonance. It seems likely that the quoted value of this width is a typographical error and should be 70(40)~eV. Indeed, including the larger value of 70~eV for $\Gamma_\gamma$ does exactly reproduce their experimental data. However, there is no experimental evidence for such a strong $\gamma$-ray decay from this state. Conversely, the $R$-matrix fit undershoots the highest cross section points in the present data. These inconsistencies emphasize the small statistical uncertainties achieved for the present measurements and those of \citet{1992ndst.book..729H} and the strong sensitivity to experimental resolution for this narrow resonance. 

Unlike most of the higher energy resonances observed in $^{14}$N+$n$, the one at $E_n$~=~433~keV has very small decay branches to both the proton or $\alpha$-particle channels, which hinders its experimental accessibility using other types of reactions. \citet{1991PhRvC..43..883W} was able to observe the corresponding resonance using the $^{14}$C$(p,n)^{14}$N reaction ($E_n$~=~434(1)~keV, $\Gamma$~=~2.4(5)~keV, $\Gamma_p$~=~9.0(15) meV), confirming a very weak proton decay branch. Finally, the very recent low-energy measurement of the $^{11}$B$(\alpha,n)^{14}$N reaction by \citet{PhysRevC.108.035809} observed the corresponding resonance in that reaction for the first time and found $\Gamma_\alpha$~=~315(48)~neV. Note that these partial widths did assume that the corresponding state had $J$~=~7/2.

\begin{figure}
    \centering
    \includegraphics[width=0.5\textwidth]{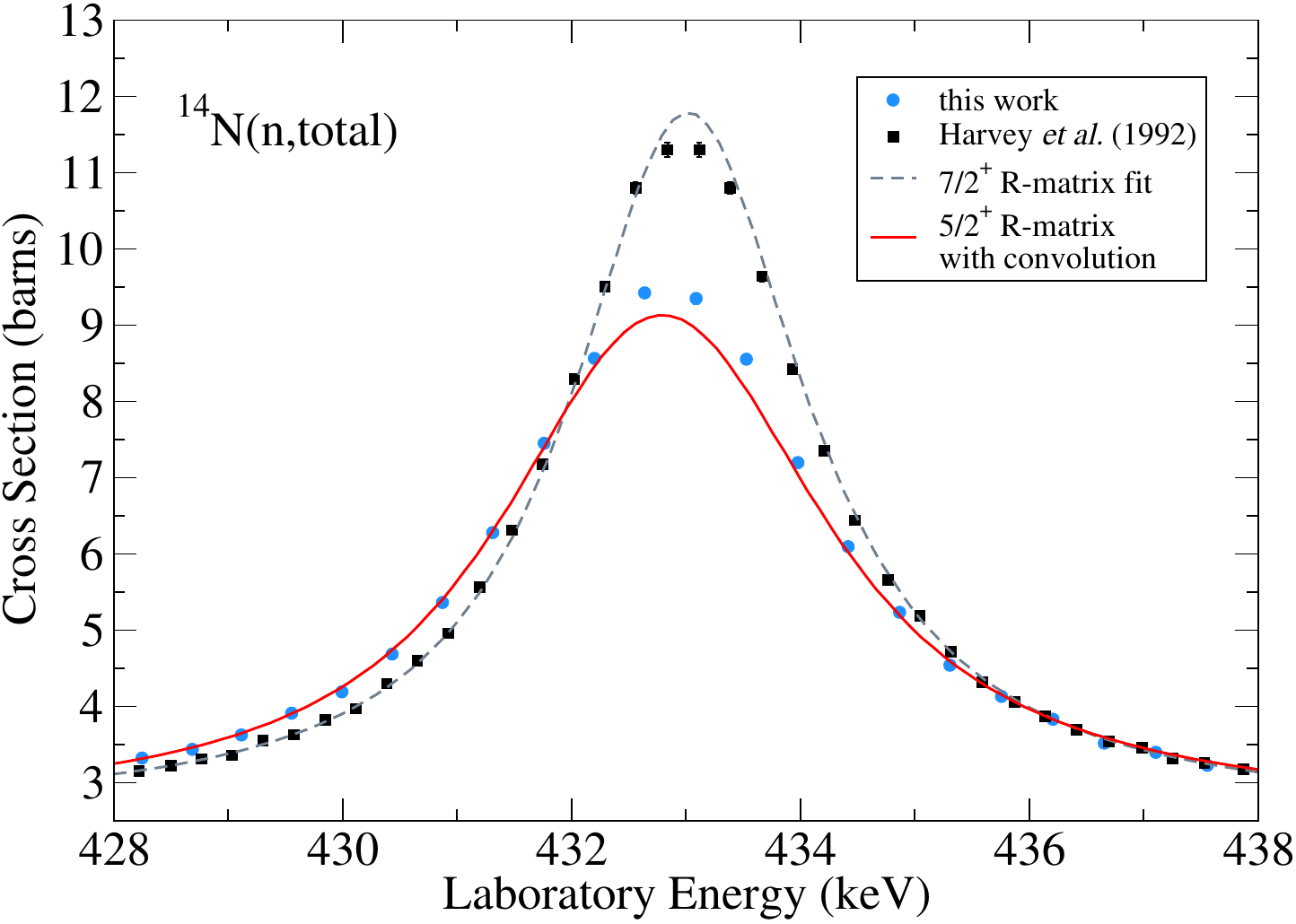}
    \caption{Comparison of the data of this work to that of \citet{1992ndst.book..729H}. The data of \citet{1992ndst.book..729H} is described well by a $R$-matrix calculation using a level of $J^\pi$~=~7/2$^+$, while that of the present work requires a lower spin of $J^\pi$~=~5/2$^+$.}
    \label{fig:433_keV_res}
\end{figure}

Fig.~\ref{fig:433_keV_res} shows a comparison between the measurement of the 433~keV resonance from this work and that of \citet{1992ndst.book..729H}. In this work, the resonance has been observed with a smaller peak cross section and a somewhat larger width. The best fit is obtained with a spin-parity of $J$~=~5/2$^+$, with $E_n$~=~433.15(1)~keV, $\Gamma_n$~=~2.82(2)~keV and $\Gamma_\alpha$~=~438(57)~neV. $\Gamma_p$ was fixed to the value of 9.0(15)~meV as reported by \citet{1991PhRvC..43..883W} and this small partial width was found to not influence the shape of the total capture cross section compared to the present uncertainties. The larger total width may seem to be an indication of an uncorrected resolution effect, but this does not seem to be the case as the nELBE resolution function has a small effect on the resonance shape at these energies. Further, the shape of the resonance is very clearly Lorentzian, and the quality of the fit worsens considerably if it is convoluted with a Gaussian function.

On the data analysis side, a constant random background is subtracted in the time-of-flight spectra, see Eq.~(\ref{eq:T}). A possible beam-related correlated background is difficult to determine in the fast neutron range, as this would require the use of ``black resonances" in the energy range of several hundred keV neutron energy~\cite{Schillebeeckx2012,ENDF333} to determine background levels for both target in/out measurements. From other transmission measurements using a 90~mm thick $^{nat}$Fe sample at nELBE \cite{Junghans2025}, the beam correlated background in resonances from 100 to 500~keV was estimated because at these resonance peaks the experimental transmission factor is determined to be 0.005 or lower. It was estimated that the beam correlated background at this energy is a factor of 2.5 higher than the uncorrelated constant background at time-of-flight longer than 4.5~$\mu$s. A straightforward subtraction of a background level that is a factor 2.5 higher than the constant background observed in the nitrogen transmission measurement both for target in/out measurements leads to a smaller transmission factor at the resonance peak (433~keV) by a factor 0.974. This corresponds to a possible higher peak cross section of 1.3\%. The observed transmission of \citet{1992ndst.book..729H} at 433~keV can be estimated to be 0.0143, which is a factor of 10 smaller than in this work (0.1557). This tends to make the observed peak value in the experiment of Harvey {\em et al.} more sensitive to the background subtraction. 

One additional complication is that there is some ambiguity in the $R$-matrix fitting of the non-resonant portion of the total neutron cross section because it is produced by some unknown number of subthreshold resonances. This allows for flexibility in the absolute total cross section resulting from the $R$-matrix fit and the experimental data also has some uncertainty in its absolute scale, which is estimated to be 1\% for the present measurements (see Sec.~\ref{sec:data_analysis}) and perhaps as small as 0.5\% for the data of \citet{1992ndst.book..729H}. If normalization factors for the data sets are allowed to vary in the fits, this does result in an improvement in the fitting of the data. For instance, for the data of \citet{1992ndst.book..729H}, an increase in the reported cross sections of 1.8\% makes it possible to fit the highest cross section data points at the peak of the 433~keV resonance, without any significant $\gamma$-ray width. Likewise, if the scale of the present data are reduced by 5\%, the fit is able to match the highest cross section data in the present measurements.

\section{\label{sec:conc} Summary}

New measurements of the $^{14}$N+$n$ total cross section have been performed at the nELBE facility at HZDR. In particular, the measurement was undertaken in order to cross-check the data reported by \citet{1992ndst.book..729H}, which have a high sensitivity and energy resolution, but they lack detailed information regarding the experimental conditions. Except for the lowest energy resonance at $E_n$~=~433~keV, the present measurements are found to be in good agreement with those reported by \citet{1992ndst.book..729H}, giving improved confidence in the present ENDF/B evaluation, which is based largely on the measurements of \citet{1992ndst.book..729H}. Further, a smaller cross section has been found for the lowest energy resonance that implies a spin-parity assignment of 5/2, rather than the value of 7/2 assigned previously. 

Because of the unparalleled sensitivity of total neutron cross sections, they provide a great deal of constraint for $R$-matrix fits that are used to evaluate resolved resonance regions. Therefore, the improved understanding and confidence in the accuracy of the $^{14}$N+$n$ total cross section of the present data and those of \citet{1992ndst.book..729H} will provide improved future evaluations for not only the $^{14}$N+$n$ total cross section but also reaction cross sections like $^{14}$N$(n,p)^{14}$C and $^{11}$B$(\alpha,n)^{14}$N. 

\begin{acknowledgments}

This research utilized resources from the Notre Dame Center for Research Computing and was supported by the National Science Foundation through Grant No. PHY-2310059 (University of Notre Dame Nuclear Science Laboratory), the Joint Institute for Nuclear Astrophysics through Grant No. PHY-1430152 (JINA Center for the Evolution of the Elements). 
A.R.J. acknowledges funding from the Euratom research and training programme 2014-2018 under grant agreement No 847594 (ARIEL).
Parts of this research were carried out at ELBE at the Helmholtz-Zentrum Dresden - Rossendorf e. V., a member of the Helmholtz Association.

\end{acknowledgments}

\bibliography{14N_ntot}

\providecommand{\noopsort}[1]{}\providecommand{\singleletter}[1]{#1}%
\begin{thebibliography}{55}%
\makeatletter
\providecommand \@ifxundefined [1]{%
 \@ifx{#1\undefined}
}%
\providecommand \@ifnum [1]{%
 \ifnum #1\expandafter \@firstoftwo
 \else \expandafter \@secondoftwo
 \fi
}%
\providecommand \@ifx [1]{%
 \ifx #1\expandafter \@firstoftwo
 \else \expandafter \@secondoftwo
 \fi
}%
\providecommand \natexlab [1]{#1}%
\providecommand \enquote  [1]{``#1''}%
\providecommand \bibnamefont  [1]{#1}%
\providecommand \bibfnamefont [1]{#1}%
\providecommand \citenamefont [1]{#1}%
\providecommand \href@noop [0]{\@secondoftwo}%
\providecommand \href [0]{\begingroup \@sanitize@url \@href}%
\providecommand \@href[1]{\@@startlink{#1}\@@href}%
\providecommand \@@href[1]{\endgroup#1\@@endlink}%
\providecommand \@sanitize@url [0]{\catcode `\\12\catcode `\$12\catcode `\&12\catcode `\#12\catcode `\^12\catcode `\_12\catcode `\%12\relax}%
\providecommand \@@startlink[1]{}%
\providecommand \@@endlink[0]{}%
\providecommand \url  [0]{\begingroup\@sanitize@url \@url }%
\providecommand \@url [1]{\endgroup\@href {#1}{\urlprefix }}%
\providecommand \urlprefix  [0]{URL }%
\providecommand \Eprint [0]{\href }%
\providecommand \doibase [0]{https://doi.org/}%
\providecommand \selectlanguage [0]{\@gobble}%
\providecommand \bibinfo  [0]{\@secondoftwo}%
\providecommand \bibfield  [0]{\@secondoftwo}%
\providecommand \translation [1]{[#1]}%
\providecommand \BibitemOpen [0]{}%
\providecommand \bibitemStop [0]{}%
\providecommand \bibitemNoStop [0]{.\EOS\space}%
\providecommand \EOS [0]{\spacefactor3000\relax}%
\providecommand \BibitemShut  [1]{\csname bibitem#1\endcsname}%
\let\auto@bib@innerbib\@empty
\bibitem [{\citenamefont {{Wallner}}\ \emph {et~al.}(2016)\citenamefont {{Wallner}}, \citenamefont {{Bichler}}, \citenamefont {{Buczak}}, \citenamefont {{Dillmann}}, \citenamefont {{K{\"a}ppeler}}, \citenamefont {{Karakas}}, \citenamefont {{Lederer}}, \citenamefont {{Lugaro}}, \citenamefont {{Mair}}, \citenamefont {{Mengoni}}, \citenamefont {{Sch{\"a}tzel}}, \citenamefont {{Steier}},\ and\ \citenamefont {{Trautvetter}}}]{2016PhRvC..93d5803W}%
  \BibitemOpen
  \bibfield  {author} {\bibinfo {author} {\bibfnamefont {A.}~\bibnamefont {{Wallner}}}, \bibinfo {author} {\bibfnamefont {M.}~\bibnamefont {{Bichler}}}, \bibinfo {author} {\bibfnamefont {K.}~\bibnamefont {{Buczak}}}, \bibinfo {author} {\bibfnamefont {I.}~\bibnamefont {{Dillmann}}}, \bibinfo {author} {\bibfnamefont {F.}~\bibnamefont {{K{\"a}ppeler}}}, \bibinfo {author} {\bibfnamefont {A.}~\bibnamefont {{Karakas}}}, \bibinfo {author} {\bibfnamefont {C.}~\bibnamefont {{Lederer}}}, \bibinfo {author} {\bibfnamefont {M.}~\bibnamefont {{Lugaro}}}, \bibinfo {author} {\bibfnamefont {K.}~\bibnamefont {{Mair}}}, \bibinfo {author} {\bibfnamefont {A.}~\bibnamefont {{Mengoni}}}, \bibinfo {author} {\bibfnamefont {G.}~\bibnamefont {{Sch{\"a}tzel}}}, \bibinfo {author} {\bibfnamefont {P.}~\bibnamefont {{Steier}}},\ and\ \bibinfo {author} {\bibfnamefont {H.~P.}\ \bibnamefont {{Trautvetter}}},\ }\href {https://doi.org/10.1103/PhysRevC.93.045803} {\bibfield  {journal} {\bibinfo  {journal} {\prc}\ }\textbf {\bibinfo {volume}
  {93}},\ \bibinfo {eid} {045803} (\bibinfo {year} {2016})}\BibitemShut {NoStop}%
\bibitem [{\citenamefont {{Koehler}}\ and\ \citenamefont {{O'Brien}}(1989)}]{1989PhRvC..39.1655K}%
  \BibitemOpen
  \bibfield  {author} {\bibinfo {author} {\bibfnamefont {P.~E.}\ \bibnamefont {{Koehler}}}\ and\ \bibinfo {author} {\bibfnamefont {H.~A.}\ \bibnamefont {{O'Brien}}},\ }\href {https://doi.org/10.1103/PhysRevC.39.1655} {\bibfield  {journal} {\bibinfo  {journal} {\prc}\ }\textbf {\bibinfo {volume} {39}},\ \bibinfo {pages} {1655} (\bibinfo {year} {1989})}\BibitemShut {NoStop}%
\bibitem [{\citenamefont {{Brehm}}\ \emph {et~al.}(1988)\citenamefont {{Brehm}}, \citenamefont {{Becker}}, \citenamefont {{Rolfs}}, \citenamefont {{Trautvetter}}, \citenamefont {{Kappeler}},\ and\ \citenamefont {{Ratynski}}}]{1988ZPhyA.330..167B}%
  \BibitemOpen
  \bibfield  {author} {\bibinfo {author} {\bibfnamefont {K.}~\bibnamefont {{Brehm}}}, \bibinfo {author} {\bibfnamefont {H.~W.}\ \bibnamefont {{Becker}}}, \bibinfo {author} {\bibfnamefont {C.}~\bibnamefont {{Rolfs}}}, \bibinfo {author} {\bibfnamefont {H.~P.}\ \bibnamefont {{Trautvetter}}}, \bibinfo {author} {\bibfnamefont {F.}~\bibnamefont {{Kappeler}}},\ and\ \bibinfo {author} {\bibfnamefont {W.}~\bibnamefont {{Ratynski}}},\ }\href {https://doi.org/10.1007/BF01293392} {\bibfield  {journal} {\bibinfo  {journal} {Zeitschrift fur Physik A Hadrons and Nuclei}\ }\textbf {\bibinfo {volume} {330}},\ \bibinfo {pages} {167} (\bibinfo {year} {1988})}\BibitemShut {NoStop}%
\bibitem [{\citenamefont {{Sanami}}\ \emph {et~al.}(1997)\citenamefont {{Sanami}}, \citenamefont {{Baba}}, \citenamefont {{Matsuyama}}, \citenamefont {{Matsuyama}}, \citenamefont {{Kiyosumi}}, \citenamefont {{Nauchi}},\ and\ \citenamefont {{Hirakawa}}}]{1997NIMPA.394..368S}%
  \BibitemOpen
  \bibfield  {author} {\bibinfo {author} {\bibfnamefont {T.}~\bibnamefont {{Sanami}}}, \bibinfo {author} {\bibfnamefont {M.}~\bibnamefont {{Baba}}}, \bibinfo {author} {\bibfnamefont {I.}~\bibnamefont {{Matsuyama}}}, \bibinfo {author} {\bibfnamefont {S.}~\bibnamefont {{Matsuyama}}}, \bibinfo {author} {\bibfnamefont {T.}~\bibnamefont {{Kiyosumi}}}, \bibinfo {author} {\bibfnamefont {Y.}~\bibnamefont {{Nauchi}}},\ and\ \bibinfo {author} {\bibfnamefont {N.}~\bibnamefont {{Hirakawa}}},\ }\href {https://doi.org/10.1016/S0168-9002(97)00698-0} {\bibfield  {journal} {\bibinfo  {journal} {Nuclear Instruments and Methods in Physics Research A}\ }\textbf {\bibinfo {volume} {394}},\ \bibinfo {pages} {368} (\bibinfo {year} {1997})}\BibitemShut {NoStop}%
\bibitem [{\citenamefont {{Shima}}\ \emph {et~al.}(1995)\citenamefont {{Shima}}, \citenamefont {{Watanabe}}, \citenamefont {{Irie}}, \citenamefont {{Sato}},\ and\ \citenamefont {{Nagai}}}]{1995NIMPA.356..347S}%
  \BibitemOpen
  \bibfield  {author} {\bibinfo {author} {\bibfnamefont {T.}~\bibnamefont {{Shima}}}, \bibinfo {author} {\bibfnamefont {K.}~\bibnamefont {{Watanabe}}}, \bibinfo {author} {\bibfnamefont {T.}~\bibnamefont {{Irie}}}, \bibinfo {author} {\bibfnamefont {H.}~\bibnamefont {{Sato}}},\ and\ \bibinfo {author} {\bibfnamefont {Y.}~\bibnamefont {{Nagai}}},\ }\href {https://doi.org/10.1016/0168-9002(94)01205-9} {\bibfield  {journal} {\bibinfo  {journal} {Nuclear Instruments and Methods in Physics Research A}\ }\textbf {\bibinfo {volume} {356}},\ \bibinfo {pages} {347} (\bibinfo {year} {1995})}\BibitemShut {NoStop}%
\bibitem [{\citenamefont {{Morgan}}(1979)}]{1979NSE....70..163M}%
  \BibitemOpen
  \bibfield  {author} {\bibinfo {author} {\bibfnamefont {G.~L.}\ \bibnamefont {{Morgan}}},\ }\href {https://doi.org/10.13182/NSE79-A19649} {\bibfield  {journal} {\bibinfo  {journal} {Nuclear Science and Engineering}\ }\textbf {\bibinfo {volume} {70}},\ \bibinfo {pages} {163} (\bibinfo {year} {1979})}\BibitemShut {NoStop}%
\bibitem [{\citenamefont {{Johnson}}\ and\ \citenamefont {{Barschall}}(1950)}]{1950PhRv...80..818J}%
  \BibitemOpen
  \bibfield  {author} {\bibinfo {author} {\bibfnamefont {C.~H.}\ \bibnamefont {{Johnson}}}\ and\ \bibinfo {author} {\bibfnamefont {H.~H.}\ \bibnamefont {{Barschall}}},\ }\href {https://doi.org/10.1103/PhysRev.80.818} {\bibfield  {journal} {\bibinfo  {journal} {Physical Review}\ }\textbf {\bibinfo {volume} {80}},\ \bibinfo {pages} {818} (\bibinfo {year} {1950})}\BibitemShut {NoStop}%
\bibitem [{\citenamefont {{Torres-S{\'a}nchez}}\ \emph {et~al.}(2023)\citenamefont {{Torres-S{\'a}nchez}}, \citenamefont {{Praena}}, \citenamefont {{Porras}}, \citenamefont {{Sabat{\'e}-Gilarte}}, \citenamefont {{Lederer-Woods}}, \citenamefont {{Aberle}}, \citenamefont {{Alcayne}}, \citenamefont {{Amaducci}}, \citenamefont {{Andrzejewski}}, \citenamefont {{Audouin}}, \citenamefont {{B{\'e}cares}}, \citenamefont {{Babiano-Suarez}}, \citenamefont {{Bacak}}, \citenamefont {{Barbagallo}}, \citenamefont {{Be{\v{c}}v{\'a}{\v{r}}}}, \citenamefont {{Bellia}}, \citenamefont {{Berthoumieux}}, \citenamefont {{Billowes}}, \citenamefont {{Bosnar}}, \citenamefont {{Brown}}, \citenamefont {{Busso}}, \citenamefont {{Caama{\~n}o}}, \citenamefont {{Caballero}}, \citenamefont {{Calvi{\~n}o}}, \citenamefont {{Calviani}}, \citenamefont {{Cano-Ott}}, \citenamefont {{Casanovas}}, \citenamefont {{Cerutti}}, \citenamefont {{Chen}}, \citenamefont {{Chiaveri}}, \citenamefont {{Colonna}}, \citenamefont {{Cort{\'e}s}}, \citenamefont
  {{Cort{\'e}s-Giraldo}}, \citenamefont {{Cosentino}}, \citenamefont {{Cristallo}}, \citenamefont {{Damone}}, \citenamefont {{Diakaki}}, \citenamefont {{Dietz}}, \citenamefont {{Domingo-Pardo}}, \citenamefont {{Dressler}}, \citenamefont {{Dupont}}, \citenamefont {{Dur{\'a}n}}, \citenamefont {{Eleme}}, \citenamefont {{Fern{\'a}ndez-Dom{\'\i}nguez}}, \citenamefont {{Ferrari}}, \citenamefont {{Ferrer}}, \citenamefont {{Finocchiaro}}, \citenamefont {{Furman}}, \citenamefont {{G{\"o}bel}}, \citenamefont {{Garg}}, \citenamefont {{Gawlik-Ramiega}}, \citenamefont {{Geslot}}, \citenamefont {{Gilardoni}}, \citenamefont {{Glodariu}}, \citenamefont {{Gon{\c{c}}alves}}, \citenamefont {{Gonz{\'a}lez-Romero}}, \citenamefont {{Guerrero}}, \citenamefont {{Gunsing}}, \citenamefont {{Harada}}, \citenamefont {{Heinitz}}, \citenamefont {{Heyse}}, \citenamefont {{Jenkins}}, \citenamefont {{Jericha}}, \citenamefont {{K{\"a}ppeler}}, \citenamefont {{Kadi}}, \citenamefont {{Kimura}}, \citenamefont {{Kivel}}, \citenamefont
  {{Kokkoris}}, \citenamefont {{Kopatch}}, \citenamefont {{Krti{\v{c}}ka}}, \citenamefont {{Kurtulgil}}, \citenamefont {{Ladarescu}}, \citenamefont {{Leeb}}, \citenamefont {{Lerendegui-Marco}}, \citenamefont {{Meo}}, \citenamefont {{Lonsdale}}, \citenamefont {{Macina}}, \citenamefont {{Manna}}, \citenamefont {{Mart{\'\i}nez}}, \citenamefont {{Masi}}, \citenamefont {{Massimi}}, \citenamefont {{Mastinu}}, \citenamefont {{Mastromarco}}, \citenamefont {{Matteucci}}, \citenamefont {{Maugeri}}, \citenamefont {{Mazzone}}, \citenamefont {{Mendoza}}, \citenamefont {{Mengoni}}, \citenamefont {{Michalopoulou}}, \citenamefont {{Milazzo}}, \citenamefont {{Mingrone}}, \citenamefont {{Musumarra}}, \citenamefont {{Negret}}, \citenamefont {{Nolte}}, \citenamefont {{Og{\'a}llar}}, \citenamefont {{Oprea}}, \citenamefont {{Patronis}}, \citenamefont {{Pavlik}}, \citenamefont {{Perkowski}}, \citenamefont {{Persanti}}, \citenamefont {{Quesada}}, \citenamefont {{Radeck}}, \citenamefont {{Ramos-Doval}}, \citenamefont {{Rauscher}},
  \citenamefont {{Reifarth}}, \citenamefont {{Rochman}}, \citenamefont {{Rubbia}}, \citenamefont {{Saxena}}, \citenamefont {{Schillebeeckx}}, \citenamefont {{Schumann}}, \citenamefont {{Smith}}, \citenamefont {{Sosnin}}, \citenamefont {{Stamatopoulos}}, \citenamefont {{Tagliente}}, \citenamefont {{Tain}}, \citenamefont {{Talip}}, \citenamefont {{Tarife{\~n}o-Saldivia}}, \citenamefont {{Tassan-Got}}, \citenamefont {{Tsinganis}}, \citenamefont {{Ulrich}}, \citenamefont {{Urlass}}, \citenamefont {{Valenta}}, \citenamefont {{Vannini}}, \citenamefont {{Variale}}, \citenamefont {{Vaz}}, \citenamefont {{Ventura}}, \citenamefont {{Vlachoudis}}, \citenamefont {{Vlastou}}, \citenamefont {{Wallner}}, \citenamefont {{Woods}}, \citenamefont {{Wright}}, \citenamefont {{{\v{Z}}ugec}},\ and\ \citenamefont {{n TOF Collaboration}}}]{2023PhRvC.107f4617T}%
  \BibitemOpen
  \bibfield  {author} {\bibinfo {author} {\bibfnamefont {P.}~\bibnamefont {{Torres-S{\'a}nchez}}}, \bibinfo {author} {\bibfnamefont {J.}~\bibnamefont {{Praena}}}, \bibinfo {author} {\bibfnamefont {I.}~\bibnamefont {{Porras}}}, \bibinfo {author} {\bibfnamefont {M.}~\bibnamefont {{Sabat{\'e}-Gilarte}}}, \bibinfo {author} {\bibfnamefont {C.}~\bibnamefont {{Lederer-Woods}}}, \bibinfo {author} {\bibfnamefont {O.}~\bibnamefont {{Aberle}}}, \bibinfo {author} {\bibfnamefont {V.}~\bibnamefont {{Alcayne}}}, \bibinfo {author} {\bibfnamefont {S.}~\bibnamefont {{Amaducci}}}, \bibinfo {author} {\bibfnamefont {J.}~\bibnamefont {{Andrzejewski}}}, \bibinfo {author} {\bibfnamefont {L.}~\bibnamefont {{Audouin}}}, \bibinfo {author} {\bibfnamefont {V.}~\bibnamefont {{B{\'e}cares}}}, \bibinfo {author} {\bibfnamefont {V.}~\bibnamefont {{Babiano-Suarez}}}, \bibinfo {author} {\bibfnamefont {M.}~\bibnamefont {{Bacak}}}, \bibinfo {author} {\bibfnamefont {M.}~\bibnamefont {{Barbagallo}}}, \bibinfo {author} {\bibfnamefont
  {F.}~\bibnamefont {{Be{\v{c}}v{\'a}{\v{r}}}}}, \bibinfo {author} {\bibfnamefont {G.}~\bibnamefont {{Bellia}}}, \bibinfo {author} {\bibfnamefont {E.}~\bibnamefont {{Berthoumieux}}}, \bibinfo {author} {\bibfnamefont {J.}~\bibnamefont {{Billowes}}}, \bibinfo {author} {\bibfnamefont {D.}~\bibnamefont {{Bosnar}}}, \bibinfo {author} {\bibfnamefont {A.}~\bibnamefont {{Brown}}}, \bibinfo {author} {\bibfnamefont {M.}~\bibnamefont {{Busso}}}, \bibinfo {author} {\bibfnamefont {M.}~\bibnamefont {{Caama{\~n}o}}}, \bibinfo {author} {\bibfnamefont {L.}~\bibnamefont {{Caballero}}}, \bibinfo {author} {\bibfnamefont {F.}~\bibnamefont {{Calvi{\~n}o}}}, \bibinfo {author} {\bibfnamefont {M.}~\bibnamefont {{Calviani}}}, \bibinfo {author} {\bibfnamefont {D.}~\bibnamefont {{Cano-Ott}}}, \bibinfo {author} {\bibfnamefont {A.}~\bibnamefont {{Casanovas}}}, \bibinfo {author} {\bibfnamefont {F.}~\bibnamefont {{Cerutti}}}, \bibinfo {author} {\bibfnamefont {Y.}~\bibnamefont {{Chen}}}, \bibinfo {author} {\bibfnamefont {E.}~\bibnamefont
  {{Chiaveri}}}, \bibinfo {author} {\bibfnamefont {N.}~\bibnamefont {{Colonna}}}, \bibinfo {author} {\bibfnamefont {G.}~\bibnamefont {{Cort{\'e}s}}}, \bibinfo {author} {\bibfnamefont {M.}~\bibnamefont {{Cort{\'e}s-Giraldo}}}, \bibinfo {author} {\bibfnamefont {L.}~\bibnamefont {{Cosentino}}}, \bibinfo {author} {\bibfnamefont {S.}~\bibnamefont {{Cristallo}}}, \bibinfo {author} {\bibfnamefont {L.-A.}\ \bibnamefont {{Damone}}}, \bibinfo {author} {\bibfnamefont {M.}~\bibnamefont {{Diakaki}}}, \bibinfo {author} {\bibfnamefont {M.}~\bibnamefont {{Dietz}}}, \bibinfo {author} {\bibfnamefont {C.}~\bibnamefont {{Domingo-Pardo}}}, \bibinfo {author} {\bibfnamefont {R.}~\bibnamefont {{Dressler}}}, \bibinfo {author} {\bibfnamefont {E.}~\bibnamefont {{Dupont}}}, \bibinfo {author} {\bibfnamefont {I.}~\bibnamefont {{Dur{\'a}n}}}, \bibinfo {author} {\bibfnamefont {Z.}~\bibnamefont {{Eleme}}}, \bibinfo {author} {\bibfnamefont {B.}~\bibnamefont {{Fern{\'a}ndez-Dom{\'\i}nguez}}}, \bibinfo {author} {\bibfnamefont {A.}~\bibnamefont
  {{Ferrari}}}, \bibinfo {author} {\bibfnamefont {F.~J.}\ \bibnamefont {{Ferrer}}}, \bibinfo {author} {\bibfnamefont {P.}~\bibnamefont {{Finocchiaro}}}, \bibinfo {author} {\bibfnamefont {V.}~\bibnamefont {{Furman}}}, \bibinfo {author} {\bibfnamefont {K.}~\bibnamefont {{G{\"o}bel}}}, \bibinfo {author} {\bibfnamefont {R.}~\bibnamefont {{Garg}}}, \bibinfo {author} {\bibfnamefont {A.}~\bibnamefont {{Gawlik-Ramiega}}}, \bibinfo {author} {\bibfnamefont {B.}~\bibnamefont {{Geslot}}}, \bibinfo {author} {\bibfnamefont {S.}~\bibnamefont {{Gilardoni}}}, \bibinfo {author} {\bibfnamefont {T.}~\bibnamefont {{Glodariu}}}, \bibinfo {author} {\bibfnamefont {I.}~\bibnamefont {{Gon{\c{c}}alves}}}, \bibinfo {author} {\bibfnamefont {E.}~\bibnamefont {{Gonz{\'a}lez-Romero}}}, \bibinfo {author} {\bibfnamefont {C.}~\bibnamefont {{Guerrero}}}, \bibinfo {author} {\bibfnamefont {F.}~\bibnamefont {{Gunsing}}}, \bibinfo {author} {\bibfnamefont {H.}~\bibnamefont {{Harada}}}, \bibinfo {author} {\bibfnamefont {S.}~\bibnamefont {{Heinitz}}},
  \bibinfo {author} {\bibfnamefont {J.}~\bibnamefont {{Heyse}}}, \bibinfo {author} {\bibfnamefont {D.}~\bibnamefont {{Jenkins}}}, \bibinfo {author} {\bibfnamefont {E.}~\bibnamefont {{Jericha}}}, \bibinfo {author} {\bibfnamefont {F.}~\bibnamefont {{K{\"a}ppeler}}}, \bibinfo {author} {\bibfnamefont {Y.}~\bibnamefont {{Kadi}}}, \bibinfo {author} {\bibfnamefont {A.}~\bibnamefont {{Kimura}}}, \bibinfo {author} {\bibfnamefont {N.}~\bibnamefont {{Kivel}}}, \bibinfo {author} {\bibfnamefont {M.}~\bibnamefont {{Kokkoris}}}, \bibinfo {author} {\bibfnamefont {Y.}~\bibnamefont {{Kopatch}}}, \bibinfo {author} {\bibfnamefont {M.}~\bibnamefont {{Krti{\v{c}}ka}}}, \bibinfo {author} {\bibfnamefont {D.}~\bibnamefont {{Kurtulgil}}}, \bibinfo {author} {\bibfnamefont {I.}~\bibnamefont {{Ladarescu}}}, \bibinfo {author} {\bibfnamefont {H.}~\bibnamefont {{Leeb}}}, \bibinfo {author} {\bibfnamefont {J.}~\bibnamefont {{Lerendegui-Marco}}}, \bibinfo {author} {\bibfnamefont {S.~L.}\ \bibnamefont {{Meo}}}, \bibinfo {author} {\bibfnamefont
  {S.-J.}\ \bibnamefont {{Lonsdale}}}, \bibinfo {author} {\bibfnamefont {D.}~\bibnamefont {{Macina}}}, \bibinfo {author} {\bibfnamefont {A.}~\bibnamefont {{Manna}}}, \bibinfo {author} {\bibfnamefont {T.}~\bibnamefont {{Mart{\'\i}nez}}}, \bibinfo {author} {\bibfnamefont {A.}~\bibnamefont {{Masi}}}, \bibinfo {author} {\bibfnamefont {C.}~\bibnamefont {{Massimi}}}, \bibinfo {author} {\bibfnamefont {P.}~\bibnamefont {{Mastinu}}}, \bibinfo {author} {\bibfnamefont {M.}~\bibnamefont {{Mastromarco}}}, \bibinfo {author} {\bibfnamefont {F.}~\bibnamefont {{Matteucci}}}, \bibinfo {author} {\bibfnamefont {E.-A.}\ \bibnamefont {{Maugeri}}}, \bibinfo {author} {\bibfnamefont {A.}~\bibnamefont {{Mazzone}}}, \bibinfo {author} {\bibfnamefont {E.}~\bibnamefont {{Mendoza}}}, \bibinfo {author} {\bibfnamefont {A.}~\bibnamefont {{Mengoni}}}, \bibinfo {author} {\bibfnamefont {V.}~\bibnamefont {{Michalopoulou}}}, \bibinfo {author} {\bibfnamefont {P.~M.}\ \bibnamefont {{Milazzo}}}, \bibinfo {author} {\bibfnamefont {F.}~\bibnamefont
  {{Mingrone}}}, \bibinfo {author} {\bibfnamefont {A.}~\bibnamefont {{Musumarra}}}, \bibinfo {author} {\bibfnamefont {A.}~\bibnamefont {{Negret}}}, \bibinfo {author} {\bibfnamefont {R.}~\bibnamefont {{Nolte}}}, \bibinfo {author} {\bibfnamefont {F.}~\bibnamefont {{Og{\'a}llar}}}, \bibinfo {author} {\bibfnamefont {A.}~\bibnamefont {{Oprea}}}, \bibinfo {author} {\bibfnamefont {N.}~\bibnamefont {{Patronis}}}, \bibinfo {author} {\bibfnamefont {A.}~\bibnamefont {{Pavlik}}}, \bibinfo {author} {\bibfnamefont {J.}~\bibnamefont {{Perkowski}}}, \bibinfo {author} {\bibfnamefont {L.}~\bibnamefont {{Persanti}}}, \bibinfo {author} {\bibfnamefont {J.-M.}\ \bibnamefont {{Quesada}}}, \bibinfo {author} {\bibfnamefont {D.}~\bibnamefont {{Radeck}}}, \bibinfo {author} {\bibfnamefont {D.}~\bibnamefont {{Ramos-Doval}}}, \bibinfo {author} {\bibfnamefont {T.}~\bibnamefont {{Rauscher}}}, \bibinfo {author} {\bibfnamefont {R.}~\bibnamefont {{Reifarth}}}, \bibinfo {author} {\bibfnamefont {D.}~\bibnamefont {{Rochman}}}, \bibinfo {author}
  {\bibfnamefont {C.}~\bibnamefont {{Rubbia}}}, \bibinfo {author} {\bibfnamefont {A.}~\bibnamefont {{Saxena}}}, \bibinfo {author} {\bibfnamefont {P.}~\bibnamefont {{Schillebeeckx}}}, \bibinfo {author} {\bibfnamefont {D.}~\bibnamefont {{Schumann}}}, \bibinfo {author} {\bibfnamefont {G.}~\bibnamefont {{Smith}}}, \bibinfo {author} {\bibfnamefont {N.}~\bibnamefont {{Sosnin}}}, \bibinfo {author} {\bibfnamefont {A.}~\bibnamefont {{Stamatopoulos}}}, \bibinfo {author} {\bibfnamefont {G.}~\bibnamefont {{Tagliente}}}, \bibinfo {author} {\bibfnamefont {J.}~\bibnamefont {{Tain}}}, \bibinfo {author} {\bibfnamefont {Z.}~\bibnamefont {{Talip}}}, \bibinfo {author} {\bibfnamefont {A.}~\bibnamefont {{Tarife{\~n}o-Saldivia}}}, \bibinfo {author} {\bibfnamefont {L.}~\bibnamefont {{Tassan-Got}}}, \bibinfo {author} {\bibfnamefont {A.}~\bibnamefont {{Tsinganis}}}, \bibinfo {author} {\bibfnamefont {J.}~\bibnamefont {{Ulrich}}}, \bibinfo {author} {\bibfnamefont {S.}~\bibnamefont {{Urlass}}}, \bibinfo {author} {\bibfnamefont
  {S.}~\bibnamefont {{Valenta}}}, \bibinfo {author} {\bibfnamefont {G.}~\bibnamefont {{Vannini}}}, \bibinfo {author} {\bibfnamefont {V.}~\bibnamefont {{Variale}}}, \bibinfo {author} {\bibfnamefont {P.}~\bibnamefont {{Vaz}}}, \bibinfo {author} {\bibfnamefont {A.}~\bibnamefont {{Ventura}}}, \bibinfo {author} {\bibfnamefont {V.}~\bibnamefont {{Vlachoudis}}}, \bibinfo {author} {\bibfnamefont {R.}~\bibnamefont {{Vlastou}}}, \bibinfo {author} {\bibfnamefont {A.}~\bibnamefont {{Wallner}}}, \bibinfo {author} {\bibfnamefont {P.}~\bibnamefont {{Woods}}}, \bibinfo {author} {\bibfnamefont {T.}~\bibnamefont {{Wright}}}, \bibinfo {author} {\bibfnamefont {P.}~\bibnamefont {{{\v{Z}}ugec}}},\ and\ \bibinfo {author} {\bibnamefont {{n TOF Collaboration}}},\ }\href {https://doi.org/10.1103/PhysRevC.107.064617} {\bibfield  {journal} {\bibinfo  {journal} {\prc}\ }\textbf {\bibinfo {volume} {107}},\ \bibinfo {eid} {064617} (\bibinfo {year} {2023})},\ \Eprint {https://arxiv.org/abs/2212.05128} {arXiv:2212.05128 [nucl-ex]}
  \BibitemShut {NoStop}%
\bibitem [{\citenamefont {Plompen}\ \emph {et~al.}(2020)\citenamefont {Plompen}, \citenamefont {Cabellos}, \citenamefont {De~Saint~Jean}, \citenamefont {Fleming}, \citenamefont {Algora}, \citenamefont {Angelone}, \citenamefont {Archier}, \citenamefont {Bauge}, \citenamefont {Bersillon}, \citenamefont {Blokhin}, \citenamefont {Cantargi}, \citenamefont {Chebboubi}, \citenamefont {Diez}, \citenamefont {Duarte}, \citenamefont {Dupont}, \citenamefont {Dyrda}, \citenamefont {Erasmus}, \citenamefont {Fiorito}, \citenamefont {Fischer}, \citenamefont {Flammini}, \citenamefont {Foligno}, \citenamefont {Gilbert}, \citenamefont {Granada}, \citenamefont {Haeck}, \citenamefont {Hambsch}, \citenamefont {Helgesson}, \citenamefont {Hilaire}, \citenamefont {Hill}, \citenamefont {Hursin}, \citenamefont {Ichou}, \citenamefont {Jacqmin}, \citenamefont {Jansky}, \citenamefont {Jouanne}, \citenamefont {Kellett}, \citenamefont {Kim}, \citenamefont {Kim}, \citenamefont {Kodeli}, \citenamefont {Koning}, \citenamefont {Konobeyev},
  \citenamefont {Kopecky}, \citenamefont {Kos}, \citenamefont {Krása}, \citenamefont {Leal}, \citenamefont {Leclaire}, \citenamefont {Leconte}, \citenamefont {Lee}, \citenamefont {Leeb}, \citenamefont {Litaize}, \citenamefont {Majerle}, \citenamefont {Márquez Damián}, \citenamefont {Michel-Sendis}, \citenamefont {Mills}, \citenamefont {Morillon}, \citenamefont {Noguère}, \citenamefont {Pecchia}, \citenamefont {Pelloni}, \citenamefont {Pereslavtsev}, \citenamefont {Perry}, \citenamefont {Rochman}, \citenamefont {Röhrmoser}, \citenamefont {Romain}, \citenamefont {Romojaro}, \citenamefont {Roubtsov}, \citenamefont {Sauvan}, \citenamefont {Schillebeeckx}, \citenamefont {Schmidt}, \citenamefont {Serot}, \citenamefont {Simakov}, \citenamefont {Sirakov}, \citenamefont {Sjöstrand}, \citenamefont {Stankovskiy}, \citenamefont {Sublet}, \citenamefont {Tamagno}, \citenamefont {Trkov}, \citenamefont {van~der Marck}, \citenamefont {Álvarez Velarde}, \citenamefont {Villari}, \citenamefont {Ware}, \citenamefont
  {Yokoyama},\ and\ \citenamefont {Žerovnik}}]{Plompen2020}%
  \BibitemOpen
  \bibfield  {author} {\bibinfo {author} {\bibfnamefont {A.~J.~M.}\ \bibnamefont {Plompen}}, \bibinfo {author} {\bibfnamefont {O.}~\bibnamefont {Cabellos}}, \bibinfo {author} {\bibfnamefont {C.}~\bibnamefont {De~Saint~Jean}}, \bibinfo {author} {\bibfnamefont {M.}~\bibnamefont {Fleming}}, \bibinfo {author} {\bibfnamefont {A.}~\bibnamefont {Algora}}, \bibinfo {author} {\bibfnamefont {M.}~\bibnamefont {Angelone}}, \bibinfo {author} {\bibfnamefont {P.}~\bibnamefont {Archier}}, \bibinfo {author} {\bibfnamefont {E.}~\bibnamefont {Bauge}}, \bibinfo {author} {\bibfnamefont {O.}~\bibnamefont {Bersillon}}, \bibinfo {author} {\bibfnamefont {A.}~\bibnamefont {Blokhin}}, \bibinfo {author} {\bibfnamefont {F.}~\bibnamefont {Cantargi}}, \bibinfo {author} {\bibfnamefont {A.}~\bibnamefont {Chebboubi}}, \bibinfo {author} {\bibfnamefont {C.}~\bibnamefont {Diez}}, \bibinfo {author} {\bibfnamefont {H.}~\bibnamefont {Duarte}}, \bibinfo {author} {\bibfnamefont {E.}~\bibnamefont {Dupont}}, \bibinfo {author} {\bibfnamefont
  {J.}~\bibnamefont {Dyrda}}, \bibinfo {author} {\bibfnamefont {B.}~\bibnamefont {Erasmus}}, \bibinfo {author} {\bibfnamefont {L.}~\bibnamefont {Fiorito}}, \bibinfo {author} {\bibfnamefont {U.}~\bibnamefont {Fischer}}, \bibinfo {author} {\bibfnamefont {D.}~\bibnamefont {Flammini}}, \bibinfo {author} {\bibfnamefont {D.}~\bibnamefont {Foligno}}, \bibinfo {author} {\bibfnamefont {M.~R.}\ \bibnamefont {Gilbert}}, \bibinfo {author} {\bibfnamefont {J.~R.}\ \bibnamefont {Granada}}, \bibinfo {author} {\bibfnamefont {W.}~\bibnamefont {Haeck}}, \bibinfo {author} {\bibfnamefont {F.-J.}\ \bibnamefont {Hambsch}}, \bibinfo {author} {\bibfnamefont {P.}~\bibnamefont {Helgesson}}, \bibinfo {author} {\bibfnamefont {S.}~\bibnamefont {Hilaire}}, \bibinfo {author} {\bibfnamefont {I.}~\bibnamefont {Hill}}, \bibinfo {author} {\bibfnamefont {M.}~\bibnamefont {Hursin}}, \bibinfo {author} {\bibfnamefont {R.}~\bibnamefont {Ichou}}, \bibinfo {author} {\bibfnamefont {R.}~\bibnamefont {Jacqmin}}, \bibinfo {author} {\bibfnamefont
  {B.}~\bibnamefont {Jansky}}, \bibinfo {author} {\bibfnamefont {C.}~\bibnamefont {Jouanne}}, \bibinfo {author} {\bibfnamefont {M.~A.}\ \bibnamefont {Kellett}}, \bibinfo {author} {\bibfnamefont {D.~H.}\ \bibnamefont {Kim}}, \bibinfo {author} {\bibfnamefont {H.~I.}\ \bibnamefont {Kim}}, \bibinfo {author} {\bibfnamefont {I.}~\bibnamefont {Kodeli}}, \bibinfo {author} {\bibfnamefont {A.~J.}\ \bibnamefont {Koning}}, \bibinfo {author} {\bibfnamefont {A.~Y.}\ \bibnamefont {Konobeyev}}, \bibinfo {author} {\bibfnamefont {S.}~\bibnamefont {Kopecky}}, \bibinfo {author} {\bibfnamefont {B.}~\bibnamefont {Kos}}, \bibinfo {author} {\bibfnamefont {A.}~\bibnamefont {Krása}}, \bibinfo {author} {\bibfnamefont {L.~C.}\ \bibnamefont {Leal}}, \bibinfo {author} {\bibfnamefont {N.}~\bibnamefont {Leclaire}}, \bibinfo {author} {\bibfnamefont {P.}~\bibnamefont {Leconte}}, \bibinfo {author} {\bibfnamefont {Y.~O.}\ \bibnamefont {Lee}}, \bibinfo {author} {\bibfnamefont {H.}~\bibnamefont {Leeb}}, \bibinfo {author} {\bibfnamefont
  {O.}~\bibnamefont {Litaize}}, \bibinfo {author} {\bibfnamefont {M.}~\bibnamefont {Majerle}}, \bibinfo {author} {\bibfnamefont {J.~I.}\ \bibnamefont {Márquez Damián}}, \bibinfo {author} {\bibfnamefont {F.}~\bibnamefont {Michel-Sendis}}, \bibinfo {author} {\bibfnamefont {R.~W.}\ \bibnamefont {Mills}}, \bibinfo {author} {\bibfnamefont {B.}~\bibnamefont {Morillon}}, \bibinfo {author} {\bibfnamefont {G.}~\bibnamefont {Noguère}}, \bibinfo {author} {\bibfnamefont {M.}~\bibnamefont {Pecchia}}, \bibinfo {author} {\bibfnamefont {S.}~\bibnamefont {Pelloni}}, \bibinfo {author} {\bibfnamefont {P.}~\bibnamefont {Pereslavtsev}}, \bibinfo {author} {\bibfnamefont {R.~J.}\ \bibnamefont {Perry}}, \bibinfo {author} {\bibfnamefont {D.}~\bibnamefont {Rochman}}, \bibinfo {author} {\bibfnamefont {A.}~\bibnamefont {Röhrmoser}}, \bibinfo {author} {\bibfnamefont {P.}~\bibnamefont {Romain}}, \bibinfo {author} {\bibfnamefont {P.}~\bibnamefont {Romojaro}}, \bibinfo {author} {\bibfnamefont {D.}~\bibnamefont {Roubtsov}}, \bibinfo
  {author} {\bibfnamefont {P.}~\bibnamefont {Sauvan}}, \bibinfo {author} {\bibfnamefont {P.}~\bibnamefont {Schillebeeckx}}, \bibinfo {author} {\bibfnamefont {K.~H.}\ \bibnamefont {Schmidt}}, \bibinfo {author} {\bibfnamefont {O.}~\bibnamefont {Serot}}, \bibinfo {author} {\bibfnamefont {S.}~\bibnamefont {Simakov}}, \bibinfo {author} {\bibfnamefont {I.}~\bibnamefont {Sirakov}}, \bibinfo {author} {\bibfnamefont {H.}~\bibnamefont {Sjöstrand}}, \bibinfo {author} {\bibfnamefont {A.}~\bibnamefont {Stankovskiy}}, \bibinfo {author} {\bibfnamefont {J.~C.}\ \bibnamefont {Sublet}}, \bibinfo {author} {\bibfnamefont {P.}~\bibnamefont {Tamagno}}, \bibinfo {author} {\bibfnamefont {A.}~\bibnamefont {Trkov}}, \bibinfo {author} {\bibfnamefont {S.}~\bibnamefont {van~der Marck}}, \bibinfo {author} {\bibfnamefont {F.}~\bibnamefont {Álvarez Velarde}}, \bibinfo {author} {\bibfnamefont {R.}~\bibnamefont {Villari}}, \bibinfo {author} {\bibfnamefont {T.~C.}\ \bibnamefont {Ware}}, \bibinfo {author} {\bibfnamefont {K.}~\bibnamefont
  {Yokoyama}},\ and\ \bibinfo {author} {\bibfnamefont {G.}~\bibnamefont {Žerovnik}},\ }\bibfield  {journal} {\bibinfo  {journal} {The European Physical Journal A}\ }\textbf {\bibinfo {volume} {56}},\ \href {https://doi.org/10.1140/epja/s10050-020-00141-9} {10.1140/epja/s10050-020-00141-9} (\bibinfo {year} {2020})\BibitemShut {NoStop}%
\bibitem [{\citenamefont {{Harvey}}\ \emph {et~al.}(1992)\citenamefont {{Harvey}}, \citenamefont {{Hill}}, \citenamefont {{Larson}},\ and\ \citenamefont {{Larson}}}]{1992ndst.book..729H}%
  \BibitemOpen
  \bibfield  {author} {\bibinfo {author} {\bibfnamefont {J.~A.}\ \bibnamefont {{Harvey}}}, \bibinfo {author} {\bibfnamefont {N.~W.}\ \bibnamefont {{Hill}}}, \bibinfo {author} {\bibfnamefont {N.~M.}\ \bibnamefont {{Larson}}},\ and\ \bibinfo {author} {\bibfnamefont {D.~C.}\ \bibnamefont {{Larson}}},\ }in\ \href {https://doi.org/10.1007/978-3-642-58113-7_208} {\emph {\bibinfo {booktitle} {Nuclear Data for Science and Technology. Series: Research Reports in Physics}}}\ (\bibinfo  {publisher} {Springer Berlin Heidelberg (Berlin, Heidelberg)},\ \bibinfo {year} {1992})\ pp.\ \bibinfo {pages} {729--731}\BibitemShut {NoStop}%
\bibitem [{\citenamefont {McLane}\ and\ \citenamefont {{Members of the Cross Section Evaluation Working Group}}(1996)}]{ENDF6_summary}%
  \BibitemOpen
  \bibfield  {author} {\bibinfo {author} {\bibfnamefont {V.}~\bibnamefont {McLane}}\ and\ \bibinfo {author} {\bibnamefont {{Members of the Cross Section Evaluation Working Group}}},\ }\href {https://www.nndc.bnl.gov/endf-b6.8/} {\emph {\bibinfo {title} {ENDF-201, ENDF/B-VI Summary Documentation Supplement I, ENDF/HE-VI Summary Documentation}}},\ \bibinfo {type} {Tech. Rep.}\ (\bibinfo  {institution} {National Nuclear Data Center, Brookhaven National Laboratory, Upton, New York 11973-5000},\ \bibinfo {year} {1996})\BibitemShut {NoStop}%
\bibitem [{\citenamefont {{Hale}}\ \emph {et~al.}(1992)\citenamefont {{Hale}}, \citenamefont {{Young}}, \citenamefont {{Chadwick}},\ and\ \citenamefont {{Chen}}}]{1992ndst.book..921H}%
  \BibitemOpen
  \bibfield  {author} {\bibinfo {author} {\bibfnamefont {G.~M.}\ \bibnamefont {{Hale}}}, \bibinfo {author} {\bibfnamefont {P.~G.}\ \bibnamefont {{Young}}}, \bibinfo {author} {\bibfnamefont {M.}~\bibnamefont {{Chadwick}}},\ and\ \bibinfo {author} {\bibfnamefont {Z.~P.}\ \bibnamefont {{Chen}}},\ }in\ \href {https://doi.org/10.1007/978-3-642-58113-7_256} {\emph {\bibinfo {booktitle} {Nuclear Data for Science and Technology. Series: Research Reports in Physics}}}\ (\bibinfo  {publisher} {Springer Berlin Heidelberg (Berlin, Heidelberg)},\ \bibinfo {year} {1992})\ pp.\ \bibinfo {pages} {921--923}\BibitemShut {NoStop}%
\bibitem [{\citenamefont {{Brown}}\ \emph {et~al.}(2018)\citenamefont {{Brown}}, \citenamefont {{Chadwick}}, \citenamefont {{Capote}}, \citenamefont {{Kahler}}, \citenamefont {{Trkov}}, \citenamefont {{Herman}}, \citenamefont {{Sonzogni}}, \citenamefont {{Danon}}, \citenamefont {{Carlson}}, \citenamefont {{Dunn}}, \citenamefont {{Smith}}, \citenamefont {{Hale}}, \citenamefont {{Arbanas}}, \citenamefont {{Arcilla}}, \citenamefont {{Bates}}, \citenamefont {{Beck}}, \citenamefont {{Becker}}, \citenamefont {{Brown}}, \citenamefont {{Casperson}}, \citenamefont {{Conlin}}, \citenamefont {{Cullen}}, \citenamefont {{Descalle}}, \citenamefont {{Firestone}}, \citenamefont {{Gaines}}, \citenamefont {{Guber}}, \citenamefont {{Hawari}}, \citenamefont {{Holmes}}, \citenamefont {{Johnson}}, \citenamefont {{Kawano}}, \citenamefont {{Kiedrowski}}, \citenamefont {{Koning}}, \citenamefont {{Kopecky}}, \citenamefont {{Leal}}, \citenamefont {{Lestone}}, \citenamefont {{Lubitz}}, \citenamefont {{M{\'a}rquez Dami{\'a}n}},
  \citenamefont {{Mattoon}}, \citenamefont {{McCutchan}}, \citenamefont {{Mughabghab}}, \citenamefont {{Navratil}}, \citenamefont {{Neudecker}}, \citenamefont {{Nobre}}, \citenamefont {{Noguere}}, \citenamefont {{Paris}}, \citenamefont {{Pigni}}, \citenamefont {{Plompen}}, \citenamefont {{Pritychenko}}, \citenamefont {{Pronyaev}}, \citenamefont {{Roubtsov}}, \citenamefont {{Rochman}}, \citenamefont {{Romano}}, \citenamefont {{Schillebeeckx}}, \citenamefont {{Simakov}}, \citenamefont {{Sin}}, \citenamefont {{Sirakov}}, \citenamefont {{Sleaford}}, \citenamefont {{Sobes}}, \citenamefont {{Soukhovitskii}}, \citenamefont {{Stetcu}}, \citenamefont {{Talou}}, \citenamefont {{Thompson}}, \citenamefont {{van der Marck}}, \citenamefont {{Welser-Sherrill}}, \citenamefont {{Wiarda}}, \citenamefont {{White}}, \citenamefont {{Wormald}}, \citenamefont {{Wright}}, \citenamefont {{Zerkle}}, \citenamefont {{{\v{Z}}erovnik}},\ and\ \citenamefont {{Zhu}}}]{2018NDS...148....1B}%
  \BibitemOpen
  \bibfield  {author} {\bibinfo {author} {\bibfnamefont {D.~A.}\ \bibnamefont {{Brown}}}, \bibinfo {author} {\bibfnamefont {M.~B.}\ \bibnamefont {{Chadwick}}}, \bibinfo {author} {\bibfnamefont {R.}~\bibnamefont {{Capote}}}, \bibinfo {author} {\bibfnamefont {A.~C.}\ \bibnamefont {{Kahler}}}, \bibinfo {author} {\bibfnamefont {A.}~\bibnamefont {{Trkov}}}, \bibinfo {author} {\bibfnamefont {M.~W.}\ \bibnamefont {{Herman}}}, \bibinfo {author} {\bibfnamefont {A.~A.}\ \bibnamefont {{Sonzogni}}}, \bibinfo {author} {\bibfnamefont {Y.}~\bibnamefont {{Danon}}}, \bibinfo {author} {\bibfnamefont {A.~D.}\ \bibnamefont {{Carlson}}}, \bibinfo {author} {\bibfnamefont {M.}~\bibnamefont {{Dunn}}}, \bibinfo {author} {\bibfnamefont {D.~L.}\ \bibnamefont {{Smith}}}, \bibinfo {author} {\bibfnamefont {G.~M.}\ \bibnamefont {{Hale}}}, \bibinfo {author} {\bibfnamefont {G.}~\bibnamefont {{Arbanas}}}, \bibinfo {author} {\bibfnamefont {R.}~\bibnamefont {{Arcilla}}}, \bibinfo {author} {\bibfnamefont {C.~R.}\ \bibnamefont {{Bates}}},
  \bibinfo {author} {\bibfnamefont {B.}~\bibnamefont {{Beck}}}, \bibinfo {author} {\bibfnamefont {B.}~\bibnamefont {{Becker}}}, \bibinfo {author} {\bibfnamefont {F.}~\bibnamefont {{Brown}}}, \bibinfo {author} {\bibfnamefont {R.~J.}\ \bibnamefont {{Casperson}}}, \bibinfo {author} {\bibfnamefont {J.}~\bibnamefont {{Conlin}}}, \bibinfo {author} {\bibfnamefont {D.~E.}\ \bibnamefont {{Cullen}}}, \bibinfo {author} {\bibfnamefont {M.~A.}\ \bibnamefont {{Descalle}}}, \bibinfo {author} {\bibfnamefont {R.}~\bibnamefont {{Firestone}}}, \bibinfo {author} {\bibfnamefont {T.}~\bibnamefont {{Gaines}}}, \bibinfo {author} {\bibfnamefont {K.~H.}\ \bibnamefont {{Guber}}}, \bibinfo {author} {\bibfnamefont {A.~I.}\ \bibnamefont {{Hawari}}}, \bibinfo {author} {\bibfnamefont {J.}~\bibnamefont {{Holmes}}}, \bibinfo {author} {\bibfnamefont {T.~D.}\ \bibnamefont {{Johnson}}}, \bibinfo {author} {\bibfnamefont {T.}~\bibnamefont {{Kawano}}}, \bibinfo {author} {\bibfnamefont {B.~C.}\ \bibnamefont {{Kiedrowski}}}, \bibinfo {author}
  {\bibfnamefont {A.~J.}\ \bibnamefont {{Koning}}}, \bibinfo {author} {\bibfnamefont {S.}~\bibnamefont {{Kopecky}}}, \bibinfo {author} {\bibfnamefont {L.}~\bibnamefont {{Leal}}}, \bibinfo {author} {\bibfnamefont {J.~P.}\ \bibnamefont {{Lestone}}}, \bibinfo {author} {\bibfnamefont {C.}~\bibnamefont {{Lubitz}}}, \bibinfo {author} {\bibfnamefont {J.~I.}\ \bibnamefont {{M{\'a}rquez Dami{\'a}n}}}, \bibinfo {author} {\bibfnamefont {C.~M.}\ \bibnamefont {{Mattoon}}}, \bibinfo {author} {\bibfnamefont {E.~A.}\ \bibnamefont {{McCutchan}}}, \bibinfo {author} {\bibfnamefont {S.}~\bibnamefont {{Mughabghab}}}, \bibinfo {author} {\bibfnamefont {P.}~\bibnamefont {{Navratil}}}, \bibinfo {author} {\bibfnamefont {D.}~\bibnamefont {{Neudecker}}}, \bibinfo {author} {\bibfnamefont {G.~P.~A.}\ \bibnamefont {{Nobre}}}, \bibinfo {author} {\bibfnamefont {G.}~\bibnamefont {{Noguere}}}, \bibinfo {author} {\bibfnamefont {M.}~\bibnamefont {{Paris}}}, \bibinfo {author} {\bibfnamefont {M.~T.}\ \bibnamefont {{Pigni}}}, \bibinfo {author}
  {\bibfnamefont {A.~J.}\ \bibnamefont {{Plompen}}}, \bibinfo {author} {\bibfnamefont {B.}~\bibnamefont {{Pritychenko}}}, \bibinfo {author} {\bibfnamefont {V.~G.}\ \bibnamefont {{Pronyaev}}}, \bibinfo {author} {\bibfnamefont {D.}~\bibnamefont {{Roubtsov}}}, \bibinfo {author} {\bibfnamefont {D.}~\bibnamefont {{Rochman}}}, \bibinfo {author} {\bibfnamefont {P.}~\bibnamefont {{Romano}}}, \bibinfo {author} {\bibfnamefont {P.}~\bibnamefont {{Schillebeeckx}}}, \bibinfo {author} {\bibfnamefont {S.}~\bibnamefont {{Simakov}}}, \bibinfo {author} {\bibfnamefont {M.}~\bibnamefont {{Sin}}}, \bibinfo {author} {\bibfnamefont {I.}~\bibnamefont {{Sirakov}}}, \bibinfo {author} {\bibfnamefont {B.}~\bibnamefont {{Sleaford}}}, \bibinfo {author} {\bibfnamefont {V.}~\bibnamefont {{Sobes}}}, \bibinfo {author} {\bibfnamefont {E.~S.}\ \bibnamefont {{Soukhovitskii}}}, \bibinfo {author} {\bibfnamefont {I.}~\bibnamefont {{Stetcu}}}, \bibinfo {author} {\bibfnamefont {P.}~\bibnamefont {{Talou}}}, \bibinfo {author} {\bibfnamefont
  {I.}~\bibnamefont {{Thompson}}}, \bibinfo {author} {\bibfnamefont {S.}~\bibnamefont {{van der Marck}}}, \bibinfo {author} {\bibfnamefont {L.}~\bibnamefont {{Welser-Sherrill}}}, \bibinfo {author} {\bibfnamefont {D.}~\bibnamefont {{Wiarda}}}, \bibinfo {author} {\bibfnamefont {M.}~\bibnamefont {{White}}}, \bibinfo {author} {\bibfnamefont {J.~L.}\ \bibnamefont {{Wormald}}}, \bibinfo {author} {\bibfnamefont {R.~Q.}\ \bibnamefont {{Wright}}}, \bibinfo {author} {\bibfnamefont {M.}~\bibnamefont {{Zerkle}}}, \bibinfo {author} {\bibfnamefont {G.}~\bibnamefont {{{\v{Z}}erovnik}}},\ and\ \bibinfo {author} {\bibfnamefont {Y.}~\bibnamefont {{Zhu}}},\ }\href {https://doi.org/10.1016/j.nds.2018.02.001} {\bibfield  {journal} {\bibinfo  {journal} {Nuclear Data Sheets}\ }\textbf {\bibinfo {volume} {148}},\ \bibinfo {pages} {1} (\bibinfo {year} {2018})}\BibitemShut {NoStop}%
\bibitem [{\citenamefont {deBoer}\ and\ \citenamefont {Dimitriou}(2018)}]{INDEN1}%
  \BibitemOpen
  \bibfield  {author} {\bibinfo {author} {\bibfnamefont {R.~J.}\ \bibnamefont {deBoer}}\ and\ \bibinfo {author} {\bibfnamefont {P.}~\bibnamefont {Dimitriou}},\ }\href {https://www-nds.iaea.org/publications/indc/indc-nds-0768/} {\emph {\bibinfo {title} {{International Nuclear Data Evaluation Network (INDEN) Meeting on the Evaluation of Light Elements}}}},\ \bibinfo {type} {Tech. Rep.}\ \bibinfo {number} {INDC(NDS)-0768}\ (\bibinfo  {institution} {International Atomic Energy Agency},\ \bibinfo {address} {Vienna, Austria},\ \bibinfo {year} {2018})\BibitemShut {NoStop}%
\bibitem [{\citenamefont {deBoer}\ and\ \citenamefont {Dimitriou}(2019)}]{INDEN2}%
  \BibitemOpen
  \bibfield  {author} {\bibinfo {author} {\bibfnamefont {R.~J.}\ \bibnamefont {deBoer}}\ and\ \bibinfo {author} {\bibfnamefont {P.}~\bibnamefont {Dimitriou}},\ }\href {https://www-nds.iaea.org/publications/indc/indc-nds-0788/} {\emph {\bibinfo {title} {{International Nuclear Data Evaluation Network (INDEN) on the Evaluation of Light Elements (2)}}}},\ \bibinfo {type} {Tech. Rep.}\ \bibinfo {number} {INDC(NDS)-0788}\ (\bibinfo  {institution} {International Atomic Energy Agency},\ \bibinfo {address} {Vienna, Austria},\ \bibinfo {year} {2019})\BibitemShut {NoStop}%
\bibitem [{\citenamefont {Leeb}\ \emph {et~al.}(2021)\citenamefont {Leeb}, \citenamefont {deBoer}, \citenamefont {Thompson},\ and\ \citenamefont {Dimitriou}}]{CM6INDEN3}%
  \BibitemOpen
  \bibfield  {author} {\bibinfo {author} {\bibfnamefont {H.}~\bibnamefont {Leeb}}, \bibinfo {author} {\bibfnamefont {R.~J.}\ \bibnamefont {deBoer}}, \bibinfo {author} {\bibfnamefont {I.}~\bibnamefont {Thompson}},\ and\ \bibinfo {author} {\bibfnamefont {P.}~\bibnamefont {Dimitriou}},\ }\href {https://www-nds.iaea.org/publications/indc/indc-nds-0827/} {\emph {\bibinfo {title} {{International Nuclear Data Evaluation Network (INDEN) on the Evaluation of Light Elements (3)}}}},\ \bibinfo {type} {Tech. Rep.}\ \bibinfo {number} {INDC(NDS)-0827}\ (\bibinfo  {institution} {International Atomic Energy Agency},\ \bibinfo {address} {Vienna, Austria},\ \bibinfo {year} {2021})\BibitemShut {NoStop}%
\bibitem [{\citenamefont {Reimer}\ \emph {et~al.}(2004)\citenamefont {Reimer}, \citenamefont {Brown},\ and\ \citenamefont {Reimer}}]{Reimer2004}%
  \BibitemOpen
  \bibfield  {author} {\bibinfo {author} {\bibfnamefont {P.~J.}\ \bibnamefont {Reimer}}, \bibinfo {author} {\bibfnamefont {T.~A.}\ \bibnamefont {Brown}},\ and\ \bibinfo {author} {\bibfnamefont {R.~W.}\ \bibnamefont {Reimer}},\ }\href {https://doi.org/10.1017/s0033822200033154} {\bibfield  {journal} {\bibinfo  {journal} {Radiocarbon}\ }\textbf {\bibinfo {volume} {46}},\ \bibinfo {pages} {1299} (\bibinfo {year} {2004})}\BibitemShut {NoStop}%
\bibitem [{\citenamefont {Burr}(2021)}]{Burr2021}%
  \BibitemOpen
  \bibfield  {author} {\bibinfo {author} {\bibfnamefont {G.~S.}\ \bibnamefont {Burr}},\ }\bibinfo {title} {Bomb carbon},\ in\ \href {https://doi.org/10.1007/978-94-007-6326-5_26-1} {\emph {\bibinfo {booktitle} {Encyclopedia of Scientific Dating Methods}}},\ \bibinfo {editor} {edited by\ \bibinfo {editor} {\bibfnamefont {W.~J.}\ \bibnamefont {Rink}}\ and\ \bibinfo {editor} {\bibfnamefont {J.}~\bibnamefont {Thompson}}}\ (\bibinfo  {publisher} {Springer Netherlands},\ \bibinfo {address} {Dordrecht},\ \bibinfo {year} {2021})\ pp.\ \bibinfo {pages} {1--1}\BibitemShut {NoStop}%
\bibitem [{\citenamefont {{Niecke}}\ \emph {et~al.}(1977)\citenamefont {{Niecke}}, \citenamefont {{Niemeier}}, \citenamefont {{Weigel}},\ and\ \citenamefont {{Wirzba-Lorenz}}}]{1977NuPhA.289..408N}%
  \BibitemOpen
  \bibfield  {author} {\bibinfo {author} {\bibfnamefont {M.}~\bibnamefont {{Niecke}}}, \bibinfo {author} {\bibfnamefont {M.}~\bibnamefont {{Niemeier}}}, \bibinfo {author} {\bibfnamefont {R.}~\bibnamefont {{Weigel}}},\ and\ \bibinfo {author} {\bibfnamefont {H.}~\bibnamefont {{Wirzba-Lorenz}}},\ }\href {https://doi.org/10.1016/0375-9474(77)90043-4} {\bibfield  {journal} {\bibinfo  {journal} {\nphysa}\ }\textbf {\bibinfo {volume} {289}},\ \bibinfo {pages} {408} (\bibinfo {year} {1977})}\BibitemShut {NoStop}%
\bibitem [{\citenamefont {{Van Der Zwan}}\ and\ \citenamefont {{Geiger}}(1975)}]{1975NuPhA.246...93V}%
  \BibitemOpen
  \bibfield  {author} {\bibinfo {author} {\bibfnamefont {L.}~\bibnamefont {{Van Der Zwan}}}\ and\ \bibinfo {author} {\bibfnamefont {K.~W.}\ \bibnamefont {{Geiger}}},\ }\href {https://doi.org/10.1016/0375-9474(75)90565-5} {\bibfield  {journal} {\bibinfo  {journal} {\nphysa}\ }\textbf {\bibinfo {volume} {246}},\ \bibinfo {pages} {93} (\bibinfo {year} {1975})}\BibitemShut {NoStop}%
\bibitem [{\citenamefont {{Wang}}\ \emph {et~al.}(1991)\citenamefont {{Wang}}, \citenamefont {{Vogelaar}},\ and\ \citenamefont {{Kavanagh}}}]{1991PhRvC..43..883W}%
  \BibitemOpen
  \bibfield  {author} {\bibinfo {author} {\bibfnamefont {T.~R.}\ \bibnamefont {{Wang}}}, \bibinfo {author} {\bibfnamefont {R.~B.}\ \bibnamefont {{Vogelaar}}},\ and\ \bibinfo {author} {\bibfnamefont {R.~W.}\ \bibnamefont {{Kavanagh}}},\ }\href {https://doi.org/10.1103/PhysRevC.43.883} {\bibfield  {journal} {\bibinfo  {journal} {\prc}\ }\textbf {\bibinfo {volume} {43}},\ \bibinfo {pages} {883} (\bibinfo {year} {1991})}\BibitemShut {NoStop}%
\bibitem [{\citenamefont {{Dayras}}\ \emph {et~al.}(1976)\citenamefont {{Dayras}}, \citenamefont {{Switkowski}},\ and\ \citenamefont {{Tombrello}}}]{1976NuPhA.261..365D}%
  \BibitemOpen
  \bibfield  {author} {\bibinfo {author} {\bibfnamefont {R.~A.}\ \bibnamefont {{Dayras}}}, \bibinfo {author} {\bibfnamefont {Z.~E.}\ \bibnamefont {{Switkowski}}},\ and\ \bibinfo {author} {\bibfnamefont {T.~A.}\ \bibnamefont {{Tombrello}}},\ }\href {https://doi.org/10.1016/0375-9474(76)90152-4} {\bibfield  {journal} {\bibinfo  {journal} {\nphysa}\ }\textbf {\bibinfo {volume} {261}},\ \bibinfo {pages} {365} (\bibinfo {year} {1976})}\BibitemShut {NoStop}%
\bibitem [{\citenamefont {{Turowiecki}}\ \emph {et~al.}(1987)\citenamefont {{Turowiecki}}, \citenamefont {{Saganek}}, \citenamefont {{Siemi{\'n}ski}}, \citenamefont {{Weso{\l}owski}},\ and\ \citenamefont {{Wilhelmi}}}]{1987NuPhA.468...29T}%
  \BibitemOpen
  \bibfield  {author} {\bibinfo {author} {\bibfnamefont {A.}~\bibnamefont {{Turowiecki}}}, \bibinfo {author} {\bibfnamefont {A.}~\bibnamefont {{Saganek}}}, \bibinfo {author} {\bibfnamefont {M.}~\bibnamefont {{Siemi{\'n}ski}}}, \bibinfo {author} {\bibfnamefont {E.}~\bibnamefont {{Weso{\l}owski}}},\ and\ \bibinfo {author} {\bibfnamefont {Z.}~\bibnamefont {{Wilhelmi}}},\ }\href {https://doi.org/10.1016/0375-9474(87)90316-2} {\bibfield  {journal} {\bibinfo  {journal} {\nphysa}\ }\textbf {\bibinfo {volume} {468}},\ \bibinfo {pages} {29} (\bibinfo {year} {1987})}\BibitemShut {NoStop}%
\bibitem [{\citenamefont {{McIntyre}}\ \emph {et~al.}(1992)\citenamefont {{McIntyre}}, \citenamefont {{Leavitt}}, \citenamefont {{Ashbaugh}}, \citenamefont {{Lin}},\ and\ \citenamefont {{Stoner}}}]{1992NIMPB..64..457M}%
  \BibitemOpen
  \bibfield  {author} {\bibinfo {author} {\bibfnamefont {L.~C.}\ \bibnamefont {{McIntyre}}}, \bibinfo {author} {\bibfnamefont {J.~A.}\ \bibnamefont {{Leavitt}}}, \bibinfo {author} {\bibfnamefont {M.~D.}\ \bibnamefont {{Ashbaugh}}}, \bibinfo {author} {\bibfnamefont {Z.}~\bibnamefont {{Lin}}},\ and\ \bibinfo {author} {\bibfnamefont {J.~O.}\ \bibnamefont {{Stoner}}},\ }\href {https://doi.org/10.1016/0168-583X(92)95515-S} {\bibfield  {journal} {\bibinfo  {journal} {Nuclear Instruments and Methods in Physics Research B}\ }\textbf {\bibinfo {volume} {64}},\ \bibinfo {pages} {457} (\bibinfo {year} {1992})}\BibitemShut {NoStop}%
\bibitem [{\citenamefont {{Liu}}\ \emph {et~al.}(1996)\citenamefont {{Liu}}, \citenamefont {{Zheng}},\ and\ \citenamefont {{Chu}}}]{1996NIMPB.108....1L}%
  \BibitemOpen
  \bibfield  {author} {\bibinfo {author} {\bibfnamefont {J.~R.}\ \bibnamefont {{Liu}}}, \bibinfo {author} {\bibfnamefont {Z.~S.}\ \bibnamefont {{Zheng}}},\ and\ \bibinfo {author} {\bibfnamefont {W.~K.}\ \bibnamefont {{Chu}}},\ }\href {https://doi.org/10.1016/0168-583X(95)00870-5} {\bibfield  {journal} {\bibinfo  {journal} {Nuclear Instruments and Methods in Physics Research B}\ }\textbf {\bibinfo {volume} {108}},\ \bibinfo {pages} {1} (\bibinfo {year} {1996})}\BibitemShut {NoStop}%
\bibitem [{\citenamefont {{Harris}}\ and\ \citenamefont {{Armstrong}}(1968)}]{1968PhRv..171.1230H}%
  \BibitemOpen
  \bibfield  {author} {\bibinfo {author} {\bibfnamefont {W.~R.}\ \bibnamefont {{Harris}}}\ and\ \bibinfo {author} {\bibfnamefont {J.~C.}\ \bibnamefont {{Armstrong}}},\ }\href {https://doi.org/10.1103/PhysRev.171.1230} {\bibfield  {journal} {\bibinfo  {journal} {Physical Review}\ }\textbf {\bibinfo {volume} {171}},\ \bibinfo {pages} {1230} (\bibinfo {year} {1968})}\BibitemShut {NoStop}%
\bibitem [{\citenamefont {{Henderson}}\ \emph {et~al.}(1968)\citenamefont {{Henderson}}, \citenamefont {{Hudspeth}},\ and\ \citenamefont {{Smith}}}]{1968PhRv..172.1058H}%
  \BibitemOpen
  \bibfield  {author} {\bibinfo {author} {\bibfnamefont {J.~D.}\ \bibnamefont {{Henderson}}}, \bibinfo {author} {\bibfnamefont {E.~L.}\ \bibnamefont {{Hudspeth}}},\ and\ \bibinfo {author} {\bibfnamefont {W.~R.}\ \bibnamefont {{Smith}}},\ }\href {https://doi.org/10.1103/PhysRev.172.1058} {\bibfield  {journal} {\bibinfo  {journal} {Physical Review}\ }\textbf {\bibinfo {volume} {172}},\ \bibinfo {pages} {1058} (\bibinfo {year} {1968})}\BibitemShut {NoStop}%
\bibitem [{\citenamefont {{Young}}\ \emph {et~al.}(1971)\citenamefont {{Young}}, \citenamefont {{Figuera}},\ and\ \citenamefont {{Steerman}}}]{1971NuPhA.173..239Y}%
  \BibitemOpen
  \bibfield  {author} {\bibinfo {author} {\bibfnamefont {F.~C.}\ \bibnamefont {{Young}}}, \bibinfo {author} {\bibfnamefont {A.~S.}\ \bibnamefont {{Figuera}}},\ and\ \bibinfo {author} {\bibfnamefont {C.~E.}\ \bibnamefont {{Steerman}}},\ }\href {https://doi.org/10.1016/0375-9474(71)90341-1} {\bibfield  {journal} {\bibinfo  {journal} {\nphysa}\ }\textbf {\bibinfo {volume} {173}},\ \bibinfo {pages} {239} (\bibinfo {year} {1971})}\BibitemShut {NoStop}%
\bibitem [{\citenamefont {{Roseborough}}\ \emph {et~al.}(1951)\citenamefont {{Roseborough}}, \citenamefont {{McCue}}, \citenamefont {{Preston}},\ and\ \citenamefont {{Goodman}}}]{1951PhRv...83.1133R}%
  \BibitemOpen
  \bibfield  {author} {\bibinfo {author} {\bibfnamefont {W.~D.}\ \bibnamefont {{Roseborough}}}, \bibinfo {author} {\bibfnamefont {J.~J.}\ \bibnamefont {{McCue}}}, \bibinfo {author} {\bibfnamefont {W.~M.}\ \bibnamefont {{Preston}}},\ and\ \bibinfo {author} {\bibfnamefont {C.}~\bibnamefont {{Goodman}}},\ }\href {https://doi.org/10.1103/PhysRev.83.1133} {\bibfield  {journal} {\bibinfo  {journal} {Physical Review}\ }\textbf {\bibinfo {volume} {83}},\ \bibinfo {pages} {1133} (\bibinfo {year} {1951})}\BibitemShut {NoStop}%
\bibitem [{\citenamefont {{Sanders}}(1956)}]{1956PhRv..104.1434S}%
  \BibitemOpen
  \bibfield  {author} {\bibinfo {author} {\bibfnamefont {R.~M.}\ \bibnamefont {{Sanders}}},\ }\href {https://doi.org/10.1103/PhysRev.104.1434} {\bibfield  {journal} {\bibinfo  {journal} {Physical Review}\ }\textbf {\bibinfo {volume} {104}},\ \bibinfo {pages} {1434} (\bibinfo {year} {1956})}\BibitemShut {NoStop}%
\bibitem [{\citenamefont {{Gibbons}}\ and\ \citenamefont {{Macklin}}(1959)}]{1959PhRv..114..571G}%
  \BibitemOpen
  \bibfield  {author} {\bibinfo {author} {\bibfnamefont {J.~H.}\ \bibnamefont {{Gibbons}}}\ and\ \bibinfo {author} {\bibfnamefont {R.~L.}\ \bibnamefont {{Macklin}}},\ }\href {https://doi.org/10.1103/PhysRev.114.571} {\bibfield  {journal} {\bibinfo  {journal} {Physical Review}\ }\textbf {\bibinfo {volume} {114}},\ \bibinfo {pages} {571} (\bibinfo {year} {1959})}\BibitemShut {NoStop}%
\bibitem [{\citenamefont {{Bartholomew}}\ \emph {et~al.}(1955)\citenamefont {{Bartholomew}}, \citenamefont {{Brown}}, \citenamefont {{Gove}}, \citenamefont {{Litherland}},\ and\ \citenamefont {{Paul}}}]{1955CaJPh..33..441B}%
  \BibitemOpen
  \bibfield  {author} {\bibinfo {author} {\bibfnamefont {G.~A.}\ \bibnamefont {{Bartholomew}}}, \bibinfo {author} {\bibfnamefont {F.}~\bibnamefont {{Brown}}}, \bibinfo {author} {\bibfnamefont {H.~E.}\ \bibnamefont {{Gove}}}, \bibinfo {author} {\bibfnamefont {A.~E.}\ \bibnamefont {{Litherland}}},\ and\ \bibinfo {author} {\bibfnamefont {E.~B.}\ \bibnamefont {{Paul}}},\ }\href {https://doi.org/10.1139/p55-053} {\bibfield  {journal} {\bibinfo  {journal} {Canadian Journal of Physics}\ }\textbf {\bibinfo {volume} {33}},\ \bibinfo {pages} {441} (\bibinfo {year} {1955})}\BibitemShut {NoStop}%
\bibitem [{\citenamefont {{Fowler}}\ and\ \citenamefont {{Johnson}}(1955)}]{1955PhRv...98..728F}%
  \BibitemOpen
  \bibfield  {author} {\bibinfo {author} {\bibfnamefont {J.~L.}\ \bibnamefont {{Fowler}}}\ and\ \bibinfo {author} {\bibfnamefont {C.~H.}\ \bibnamefont {{Johnson}}},\ }\href {https://doi.org/10.1103/PhysRev.98.728} {\bibfield  {journal} {\bibinfo  {journal} {Physical Review}\ }\textbf {\bibinfo {volume} {98}},\ \bibinfo {pages} {728} (\bibinfo {year} {1955})}\BibitemShut {NoStop}%
\bibitem [{\citenamefont {{Gabbard}}\ \emph {et~al.}(1959)\citenamefont {{Gabbard}}, \citenamefont {{Bichsel}},\ and\ \citenamefont {{Bonner}}}]{1959NucPh..14..277G}%
  \BibitemOpen
  \bibfield  {author} {\bibinfo {author} {\bibfnamefont {F.}~\bibnamefont {{Gabbard}}}, \bibinfo {author} {\bibfnamefont {H.}~\bibnamefont {{Bichsel}}},\ and\ \bibinfo {author} {\bibfnamefont {T.~W.}\ \bibnamefont {{Bonner}}},\ }\href {https://doi.org/10.1016/0029-5582(59)90013-6} {\bibfield  {journal} {\bibinfo  {journal} {Nuclear Physics}\ }\textbf {\bibinfo {volume} {14}},\ \bibinfo {pages} {277} (\bibinfo {year} {1959})}\BibitemShut {NoStop}%
\bibitem [{\citenamefont {{Hinchey}}\ \emph {et~al.}(1952)\citenamefont {{Hinchey}}, \citenamefont {{Stelson}},\ and\ \citenamefont {{Preston}}}]{1952PhRv...86..483H}%
  \BibitemOpen
  \bibfield  {author} {\bibinfo {author} {\bibfnamefont {J.~J.}\ \bibnamefont {{Hinchey}}}, \bibinfo {author} {\bibfnamefont {P.~H.}\ \bibnamefont {{Stelson}}},\ and\ \bibinfo {author} {\bibfnamefont {W.~M.}\ \bibnamefont {{Preston}}},\ }\href {https://doi.org/10.1103/PhysRev.86.483} {\bibfield  {journal} {\bibinfo  {journal} {Physical Review}\ }\textbf {\bibinfo {volume} {86}},\ \bibinfo {pages} {483} (\bibinfo {year} {1952})}\BibitemShut {NoStop}%
\bibitem [{\citenamefont {{Johnson}}\ \emph {et~al.}(1951)\citenamefont {{Johnson}}, \citenamefont {{Petree}},\ and\ \citenamefont {{Adair}}}]{1951PhRv...84..775J}%
  \BibitemOpen
  \bibfield  {author} {\bibinfo {author} {\bibfnamefont {C.~H.}\ \bibnamefont {{Johnson}}}, \bibinfo {author} {\bibfnamefont {B.}~\bibnamefont {{Petree}}},\ and\ \bibinfo {author} {\bibfnamefont {R.~K.}\ \bibnamefont {{Adair}}},\ }\href {https://doi.org/10.1103/PhysRev.84.775} {\bibfield  {journal} {\bibinfo  {journal} {Physical Review}\ }\textbf {\bibinfo {volume} {84}},\ \bibinfo {pages} {775} (\bibinfo {year} {1951})}\BibitemShut {NoStop}%
\bibitem [{\citenamefont {Klug}\ \emph {et~al.}(2007)\citenamefont {Klug}, \citenamefont {Altstadt}, \citenamefont {Beckert}, \citenamefont {Beyer}, \citenamefont {Freiesleben}, \citenamefont {Galindo}, \citenamefont {Grosse}, \citenamefont {Junghans}, \citenamefont {L{\'{e}}gr{\'{a}}dy}, \citenamefont {Naumann}, \citenamefont {Noack}, \citenamefont {Rusev}, \citenamefont {Schilling}, \citenamefont {Schlenk}, \citenamefont {Schneider}, \citenamefont {Wagner},\ and\ \citenamefont {Weiss}}]{Klug2007}%
  \BibitemOpen
  \bibfield  {author} {\bibinfo {author} {\bibfnamefont {J.}~\bibnamefont {Klug}}, \bibinfo {author} {\bibfnamefont {E.}~\bibnamefont {Altstadt}}, \bibinfo {author} {\bibfnamefont {C.}~\bibnamefont {Beckert}}, \bibinfo {author} {\bibfnamefont {R.}~\bibnamefont {Beyer}}, \bibinfo {author} {\bibfnamefont {H.}~\bibnamefont {Freiesleben}}, \bibinfo {author} {\bibfnamefont {V.}~\bibnamefont {Galindo}}, \bibinfo {author} {\bibfnamefont {E.}~\bibnamefont {Grosse}}, \bibinfo {author} {\bibfnamefont {A.}~\bibnamefont {Junghans}}, \bibinfo {author} {\bibfnamefont {D.}~\bibnamefont {L{\'{e}}gr{\'{a}}dy}}, \bibinfo {author} {\bibfnamefont {B.}~\bibnamefont {Naumann}}, \bibinfo {author} {\bibfnamefont {K.}~\bibnamefont {Noack}}, \bibinfo {author} {\bibfnamefont {G.}~\bibnamefont {Rusev}}, \bibinfo {author} {\bibfnamefont {K.}~\bibnamefont {Schilling}}, \bibinfo {author} {\bibfnamefont {R.}~\bibnamefont {Schlenk}}, \bibinfo {author} {\bibfnamefont {S.}~\bibnamefont {Schneider}}, \bibinfo {author} {\bibfnamefont
  {A.}~\bibnamefont {Wagner}},\ and\ \bibinfo {author} {\bibfnamefont {F.-P.}\ \bibnamefont {Weiss}},\ }\href {https://doi.org/10.1016/j.nima.2007.04.132} {\bibfield  {journal} {\bibinfo  {journal} {Nuclear Instruments and Methods in Physics Research Section A: Accelerators, Spectrometers, Detectors and Associated Equipment}\ }\textbf {\bibinfo {volume} {577}},\ \bibinfo {pages} {641} (\bibinfo {year} {2007})}\BibitemShut {NoStop}%
\bibitem [{\citenamefont {Beyer}\ \emph {et~al.}(2013)\citenamefont {Beyer}, \citenamefont {Birgersson}, \citenamefont {Elekes}, \citenamefont {Ferrari}, \citenamefont {Grosse}, \citenamefont {Hannaske}, \citenamefont {Junghans}, \citenamefont {Kögler}, \citenamefont {Massarczyk}, \citenamefont {Mati{\'{c}}}, \citenamefont {Nolte}, \citenamefont {Schwengner},\ and\ \citenamefont {Wagner}}]{Beyer2013}%
  \BibitemOpen
  \bibfield  {author} {\bibinfo {author} {\bibfnamefont {R.}~\bibnamefont {Beyer}}, \bibinfo {author} {\bibfnamefont {E.}~\bibnamefont {Birgersson}}, \bibinfo {author} {\bibfnamefont {Z.}~\bibnamefont {Elekes}}, \bibinfo {author} {\bibfnamefont {A.}~\bibnamefont {Ferrari}}, \bibinfo {author} {\bibfnamefont {E.}~\bibnamefont {Grosse}}, \bibinfo {author} {\bibfnamefont {R.}~\bibnamefont {Hannaske}}, \bibinfo {author} {\bibfnamefont {A.}~\bibnamefont {Junghans}}, \bibinfo {author} {\bibfnamefont {T.}~\bibnamefont {Kögler}}, \bibinfo {author} {\bibfnamefont {R.}~\bibnamefont {Massarczyk}}, \bibinfo {author} {\bibfnamefont {A.}~\bibnamefont {Mati{\'{c}}}}, \bibinfo {author} {\bibfnamefont {R.}~\bibnamefont {Nolte}}, \bibinfo {author} {\bibfnamefont {R.}~\bibnamefont {Schwengner}},\ and\ \bibinfo {author} {\bibfnamefont {A.}~\bibnamefont {Wagner}},\ }\href {https://doi.org/10.1016/j.nima.2013.05.010} {\bibfield  {journal} {\bibinfo  {journal} {Nuclear Instruments and Methods in Physics Research Section A:
  Accelerators, Spectrometers, Detectors and Associated Equipment}\ }\textbf {\bibinfo {volume} {723}},\ \bibinfo {pages} {151} (\bibinfo {year} {2013})}\BibitemShut {NoStop}%
\bibitem [{\citenamefont {Beyer}\ \emph {et~al.}(2007)\citenamefont {Beyer}, \citenamefont {Grosse}, \citenamefont {Heidel}, \citenamefont {Hutsch}, \citenamefont {Junghans}, \citenamefont {Klug}, \citenamefont {L{\'{e}}gr{\'{a}}dy}, \citenamefont {Nolte}, \citenamefont {Röttger}, \citenamefont {Sobiella},\ and\ \citenamefont {Wagner}}]{Beyer2007}%
  \BibitemOpen
  \bibfield  {author} {\bibinfo {author} {\bibfnamefont {R.}~\bibnamefont {Beyer}}, \bibinfo {author} {\bibfnamefont {E.}~\bibnamefont {Grosse}}, \bibinfo {author} {\bibfnamefont {K.}~\bibnamefont {Heidel}}, \bibinfo {author} {\bibfnamefont {J.}~\bibnamefont {Hutsch}}, \bibinfo {author} {\bibfnamefont {A.}~\bibnamefont {Junghans}}, \bibinfo {author} {\bibfnamefont {J.}~\bibnamefont {Klug}}, \bibinfo {author} {\bibfnamefont {D.}~\bibnamefont {L{\'{e}}gr{\'{a}}dy}}, \bibinfo {author} {\bibfnamefont {R.}~\bibnamefont {Nolte}}, \bibinfo {author} {\bibfnamefont {S.}~\bibnamefont {Röttger}}, \bibinfo {author} {\bibfnamefont {M.}~\bibnamefont {Sobiella}},\ and\ \bibinfo {author} {\bibfnamefont {A.}~\bibnamefont {Wagner}},\ }\href {https://doi.org/10.1016/j.nima.2007.02.096} {\bibfield  {journal} {\bibinfo  {journal} {Nuclear Instruments and Methods in Physics Research Section A: Accelerators, Spectrometers, Detectors and Associated Equipment}\ }\textbf {\bibinfo {volume} {575}},\ \bibinfo {pages} {449} (\bibinfo
  {year} {2007})}\BibitemShut {NoStop}%
\bibitem [{\citenamefont {Mughabghab}(2018)}]{Mughabghab2018}%
  \BibitemOpen
  \bibfield  {author} {\bibinfo {author} {\bibfnamefont {S.}~\bibnamefont {Mughabghab}},\ }\href {https://doi.org/10.1016/c2015-0-00522-6} {\emph {\bibinfo {title} {Atlas of Neutron Resonances}}}\ (\bibinfo  {publisher} {Elsevier},\ \bibinfo {year} {2018})\BibitemShut {NoStop}%
\bibitem [{\citenamefont {Lemmon}\ \emph {et~al.}(2016)\citenamefont {Lemmon}, \citenamefont {McLinden},\ and\ \citenamefont {Friend}}]{Linstrom2016}%
  \BibitemOpen
  \bibfield  {author} {\bibinfo {author} {\bibfnamefont {E.~W.}\ \bibnamefont {Lemmon}}, \bibinfo {author} {\bibfnamefont {M.~O.}\ \bibnamefont {McLinden}},\ and\ \bibinfo {author} {\bibfnamefont {D.~G.}\ \bibnamefont {Friend}},\ }\href {https://webbook.nist.gov/chemistry/} {\emph {\bibinfo {title} {Thermophysical Properties of Fluid Systems}}},\ edited by\ \bibinfo {editor} {\bibfnamefont {P.}~\bibnamefont {Linstrom}}\ and\ \bibinfo {editor} {\bibfnamefont {W.}~\bibnamefont {Mallard}},\ Vol.\ \bibinfo {volume} {NIST Chemistry WebBook, NIST Standard Reference Database Number 69,}\ (\bibinfo  {publisher} {National Institute of Standards and Technology, Gaithersburg MD, 20899},\ \bibinfo {year} {2016})\BibitemShut {NoStop}%
\bibitem [{MBS()}]{MBS}%
  \BibitemOpen
  \href@noop {} {\bibinfo {title} {{Multi Branch System (MBS)}}},\ \bibinfo {howpublished} {\url{https://www.gsi.de/en/work/research/experiment_electronics/data_processing/data_acquisition/mbs}},\ \bibinfo {note} {accessed: 2025-03-17}\BibitemShut {NoStop}%
\bibitem [{\citenamefont {Beyer}\ \emph {et~al.}(2018)\citenamefont {Beyer}, \citenamefont {Junghans}, \citenamefont {Schillebeeckx}, \citenamefont {Sirakov}, \citenamefont {Song}, \citenamefont {Bemmerer}, \citenamefont {Capote}, \citenamefont {Ferrari}, \citenamefont {Hartmann}, \citenamefont {Hannaske}, \citenamefont {Heyse}, \citenamefont {Kim}, \citenamefont {Kim}, \citenamefont {Kögler}, \citenamefont {Lee}, \citenamefont {Lee}, \citenamefont {Massarczyk}, \citenamefont {Müller}, \citenamefont {Reinhardt}, \citenamefont {Röder}, \citenamefont {Schmidt}, \citenamefont {Schwengner}, \citenamefont {Szücs}, \citenamefont {Tak{\'{a}}cs}, \citenamefont {Wagner}, \citenamefont {Wagner},\ and\ \citenamefont {Yang}}]{Beyer2018}%
  \BibitemOpen
  \bibfield  {author} {\bibinfo {author} {\bibfnamefont {R.}~\bibnamefont {Beyer}}, \bibinfo {author} {\bibfnamefont {A.~R.}\ \bibnamefont {Junghans}}, \bibinfo {author} {\bibfnamefont {P.}~\bibnamefont {Schillebeeckx}}, \bibinfo {author} {\bibfnamefont {I.}~\bibnamefont {Sirakov}}, \bibinfo {author} {\bibfnamefont {T.-Y.}\ \bibnamefont {Song}}, \bibinfo {author} {\bibfnamefont {D.}~\bibnamefont {Bemmerer}}, \bibinfo {author} {\bibfnamefont {R.}~\bibnamefont {Capote}}, \bibinfo {author} {\bibfnamefont {A.}~\bibnamefont {Ferrari}}, \bibinfo {author} {\bibfnamefont {A.}~\bibnamefont {Hartmann}}, \bibinfo {author} {\bibfnamefont {R.}~\bibnamefont {Hannaske}}, \bibinfo {author} {\bibfnamefont {J.}~\bibnamefont {Heyse}}, \bibinfo {author} {\bibfnamefont {H.~I.}\ \bibnamefont {Kim}}, \bibinfo {author} {\bibfnamefont {J.~W.}\ \bibnamefont {Kim}}, \bibinfo {author} {\bibfnamefont {T.}~\bibnamefont {Kögler}}, \bibinfo {author} {\bibfnamefont {C.~W.}\ \bibnamefont {Lee}}, \bibinfo {author} {\bibfnamefont {Y.-O.}\
  \bibnamefont {Lee}}, \bibinfo {author} {\bibfnamefont {R.}~\bibnamefont {Massarczyk}}, \bibinfo {author} {\bibfnamefont {S.~E.}\ \bibnamefont {Müller}}, \bibinfo {author} {\bibfnamefont {T.~P.}\ \bibnamefont {Reinhardt}}, \bibinfo {author} {\bibfnamefont {M.}~\bibnamefont {Röder}}, \bibinfo {author} {\bibfnamefont {K.}~\bibnamefont {Schmidt}}, \bibinfo {author} {\bibfnamefont {R.}~\bibnamefont {Schwengner}}, \bibinfo {author} {\bibfnamefont {T.}~\bibnamefont {Szücs}}, \bibinfo {author} {\bibfnamefont {M.~P.}\ \bibnamefont {Tak{\'{a}}cs}}, \bibinfo {author} {\bibfnamefont {A.}~\bibnamefont {Wagner}}, \bibinfo {author} {\bibfnamefont {L.}~\bibnamefont {Wagner}},\ and\ \bibinfo {author} {\bibfnamefont {S.-C.}\ \bibnamefont {Yang}},\ }\href {https://doi.org/10.1140/epja/i2018-12505-7} {\bibfield  {journal} {\bibinfo  {journal} {The European Physical Journal A}\ }\textbf {\bibinfo {volume} {54}},\ \bibinfo {pages} {81} (\bibinfo {year} {2018})}\BibitemShut {NoStop}%
\bibitem [{\citenamefont {Hannaske}\ \emph {et~al.}(2013)\citenamefont {Hannaske}, \citenamefont {Elekes}, \citenamefont {Beyer}, \citenamefont {Junghans}, \citenamefont {Bemmerer}, \citenamefont {Birgersson}, \citenamefont {Ferrari}, \citenamefont {Grosse}, \citenamefont {Kempe}, \citenamefont {Kögler}, \citenamefont {Marta}, \citenamefont {Massarczyk}, \citenamefont {Matic}, \citenamefont {Schramm}, \citenamefont {Schwengner},\ and\ \citenamefont {Wagner}}]{Hannaske2013}%
  \BibitemOpen
  \bibfield  {author} {\bibinfo {author} {\bibfnamefont {R.}~\bibnamefont {Hannaske}}, \bibinfo {author} {\bibfnamefont {Z.}~\bibnamefont {Elekes}}, \bibinfo {author} {\bibfnamefont {R.}~\bibnamefont {Beyer}}, \bibinfo {author} {\bibfnamefont {A.}~\bibnamefont {Junghans}}, \bibinfo {author} {\bibfnamefont {D.}~\bibnamefont {Bemmerer}}, \bibinfo {author} {\bibfnamefont {E.}~\bibnamefont {Birgersson}}, \bibinfo {author} {\bibfnamefont {A.}~\bibnamefont {Ferrari}}, \bibinfo {author} {\bibfnamefont {E.}~\bibnamefont {Grosse}}, \bibinfo {author} {\bibfnamefont {M.}~\bibnamefont {Kempe}}, \bibinfo {author} {\bibfnamefont {T.}~\bibnamefont {Kögler}}, \bibinfo {author} {\bibfnamefont {M.}~\bibnamefont {Marta}}, \bibinfo {author} {\bibfnamefont {R.}~\bibnamefont {Massarczyk}}, \bibinfo {author} {\bibfnamefont {A.}~\bibnamefont {Matic}}, \bibinfo {author} {\bibfnamefont {G.}~\bibnamefont {Schramm}}, \bibinfo {author} {\bibfnamefont {R.}~\bibnamefont {Schwengner}},\ and\ \bibinfo {author} {\bibfnamefont
  {A.}~\bibnamefont {Wagner}},\ }\href {https://doi.org/10.1140/epja/i2013-13137-1} {\bibfield  {journal} {\bibinfo  {journal} {The European Physical Journal A}\ }\textbf {\bibinfo {volume} {49}},\ \bibinfo {pages} {137} (\bibinfo {year} {2013})}\BibitemShut {NoStop}%
\bibitem [{\citenamefont {{Azuma}}\ \emph {et~al.}(2010)\citenamefont {{Azuma}}, \citenamefont {{Uberseder}}, \citenamefont {{Simpson}}, \citenamefont {{Brune}}, \citenamefont {{Costantini}}, \citenamefont {{de Boer}}, \citenamefont {{G{\"o}rres}}, \citenamefont {{Heil}}, \citenamefont {{Leblanc}}, \citenamefont {{Ugalde}},\ and\ \citenamefont {{Wiescher}}}]{2010PhRvC..81d5805A}%
  \BibitemOpen
  \bibfield  {author} {\bibinfo {author} {\bibfnamefont {R.~E.}\ \bibnamefont {{Azuma}}}, \bibinfo {author} {\bibfnamefont {E.}~\bibnamefont {{Uberseder}}}, \bibinfo {author} {\bibfnamefont {E.~C.}\ \bibnamefont {{Simpson}}}, \bibinfo {author} {\bibfnamefont {C.~R.}\ \bibnamefont {{Brune}}}, \bibinfo {author} {\bibfnamefont {H.}~\bibnamefont {{Costantini}}}, \bibinfo {author} {\bibfnamefont {R.~J.}\ \bibnamefont {{de Boer}}}, \bibinfo {author} {\bibfnamefont {J.}~\bibnamefont {{G{\"o}rres}}}, \bibinfo {author} {\bibfnamefont {M.}~\bibnamefont {{Heil}}}, \bibinfo {author} {\bibfnamefont {P.~J.}\ \bibnamefont {{Leblanc}}}, \bibinfo {author} {\bibfnamefont {C.}~\bibnamefont {{Ugalde}}},\ and\ \bibinfo {author} {\bibfnamefont {M.}~\bibnamefont {{Wiescher}}},\ }\href {https://doi.org/10.1103/PhysRevC.81.045805} {\bibfield  {journal} {\bibinfo  {journal} {\prc}\ }\textbf {\bibinfo {volume} {81}},\ \bibinfo {eid} {045805} (\bibinfo {year} {2010})}\BibitemShut {NoStop}%
\bibitem [{\citenamefont {{deBoer}}\ \emph {et~al.}(2020)\citenamefont {{deBoer}}, \citenamefont {{Liu}}, \citenamefont {{Chen}}, \citenamefont {{Couder}}, \citenamefont {{G{\"o}rres}}, \citenamefont {{Lamere}}, \citenamefont {{Long}}, \citenamefont {{Lyons}}, \citenamefont {{Manukyan}}, \citenamefont {{Morales}}, \citenamefont {{Robertson}}, \citenamefont {{Seymour}}, \citenamefont {{Seymour}}, \citenamefont {{Stech}}, \citenamefont {{Vande Kolk}},\ and\ \citenamefont {{Wiescher}}}]{2020JPhCS1668a2011D}%
  \BibitemOpen
  \bibfield  {author} {\bibinfo {author} {\bibfnamefont {R.~J.}\ \bibnamefont {{deBoer}}}, \bibinfo {author} {\bibfnamefont {Q.}~\bibnamefont {{Liu}}}, \bibinfo {author} {\bibfnamefont {Y.}~\bibnamefont {{Chen}}}, \bibinfo {author} {\bibfnamefont {M.}~\bibnamefont {{Couder}}}, \bibinfo {author} {\bibfnamefont {J.}~\bibnamefont {{G{\"o}rres}}}, \bibinfo {author} {\bibfnamefont {E.}~\bibnamefont {{Lamere}}}, \bibinfo {author} {\bibfnamefont {A.~M.}\ \bibnamefont {{Long}}}, \bibinfo {author} {\bibfnamefont {S.}~\bibnamefont {{Lyons}}}, \bibinfo {author} {\bibfnamefont {K.}~\bibnamefont {{Manukyan}}}, \bibinfo {author} {\bibfnamefont {L.}~\bibnamefont {{Morales}}}, \bibinfo {author} {\bibfnamefont {D.}~\bibnamefont {{Robertson}}}, \bibinfo {author} {\bibfnamefont {C.}~\bibnamefont {{Seymour}}}, \bibinfo {author} {\bibfnamefont {G.}~\bibnamefont {{Seymour}}}, \bibinfo {author} {\bibfnamefont {E.}~\bibnamefont {{Stech}}}, \bibinfo {author} {\bibfnamefont {B.}~\bibnamefont {{Vande Kolk}}},\ and\ \bibinfo {author}
  {\bibfnamefont {M.}~\bibnamefont {{Wiescher}}},\ }in\ \href {https://doi.org/10.1088/1742-6596/1668/1/012011} {\emph {\bibinfo {booktitle} {Journal of Physics Conference Series}}},\ \bibinfo {series} {Journal of Physics Conference Series}, Vol.\ \bibinfo {volume} {1668}\ (\bibinfo {year} {2020})\ p.\ \bibinfo {pages} {012011}\BibitemShut {NoStop}%
\bibitem [{\citenamefont {Borgwardt}\ \emph {et~al.}(2023)\citenamefont {Borgwardt}, \citenamefont {deBoer}, \citenamefont {Boeltzig}, \citenamefont {Couder}, \citenamefont {G\"orres}, \citenamefont {Gula}, \citenamefont {Hanhardt}, \citenamefont {Manukyan}, \citenamefont {Kadlecek}, \citenamefont {Robertson}, \citenamefont {Strieder},\ and\ \citenamefont {Wiescher}}]{PhysRevC.108.035809}%
  \BibitemOpen
  \bibfield  {author} {\bibinfo {author} {\bibfnamefont {T.~C.}\ \bibnamefont {Borgwardt}}, \bibinfo {author} {\bibfnamefont {R.~J.}\ \bibnamefont {deBoer}}, \bibinfo {author} {\bibfnamefont {A.}~\bibnamefont {Boeltzig}}, \bibinfo {author} {\bibfnamefont {M.}~\bibnamefont {Couder}}, \bibinfo {author} {\bibfnamefont {J.}~\bibnamefont {G\"orres}}, \bibinfo {author} {\bibfnamefont {A.}~\bibnamefont {Gula}}, \bibinfo {author} {\bibfnamefont {M.}~\bibnamefont {Hanhardt}}, \bibinfo {author} {\bibfnamefont {K.~V.}\ \bibnamefont {Manukyan}}, \bibinfo {author} {\bibfnamefont {T.}~\bibnamefont {Kadlecek}}, \bibinfo {author} {\bibfnamefont {D.}~\bibnamefont {Robertson}}, \bibinfo {author} {\bibfnamefont {F.}~\bibnamefont {Strieder}},\ and\ \bibinfo {author} {\bibfnamefont {M.}~\bibnamefont {Wiescher}},\ }\href {https://doi.org/10.1103/PhysRevC.108.035809} {\bibfield  {journal} {\bibinfo  {journal} {Phys. Rev. C}\ }\textbf {\bibinfo {volume} {108}},\ \bibinfo {pages} {035809} (\bibinfo {year} {2023})}\BibitemShut
  {NoStop}%
\bibitem [{\citenamefont {{Brune}}(2002)}]{2002PhRvC..66d4611B}%
  \BibitemOpen
  \bibfield  {author} {\bibinfo {author} {\bibfnamefont {C.~R.}\ \bibnamefont {{Brune}}},\ }\href {https://doi.org/10.1103/PhysRevC.66.044611} {\bibfield  {journal} {\bibinfo  {journal} {\prc}\ }\textbf {\bibinfo {volume} {66}},\ \bibinfo {eid} {044611} (\bibinfo {year} {2002})},\ \Eprint {https://arxiv.org/abs/nucl-th/0207048} {arXiv:nucl-th/0207048 [nucl-th]} \BibitemShut {NoStop}%
\bibitem [{\citenamefont {{Lane}}\ and\ \citenamefont {{Thomas}}(1958)}]{1958RvMP...30..257L}%
  \BibitemOpen
  \bibfield  {author} {\bibinfo {author} {\bibfnamefont {A.~M.}\ \bibnamefont {{Lane}}}\ and\ \bibinfo {author} {\bibfnamefont {R.~G.}\ \bibnamefont {{Thomas}}},\ }\href {https://doi.org/10.1103/RevModPhys.30.257} {\bibfield  {journal} {\bibinfo  {journal} {Reviews of Modern Physics}\ }\textbf {\bibinfo {volume} {30}},\ \bibinfo {pages} {257} (\bibinfo {year} {1958})}\BibitemShut {NoStop}%
\bibitem [{\citenamefont {{Odell}}\ \emph {et~al.}(2022)\citenamefont {{Odell}}, \citenamefont {{Brune}}, \citenamefont {{Phillips}}, \citenamefont {{deBoer}},\ and\ \citenamefont {{Paneru}}}]{2022FrP....10.8476O}%
  \BibitemOpen
  \bibfield  {author} {\bibinfo {author} {\bibfnamefont {D.}~\bibnamefont {{Odell}}}, \bibinfo {author} {\bibfnamefont {C.~R.}\ \bibnamefont {{Brune}}}, \bibinfo {author} {\bibfnamefont {D.~R.}\ \bibnamefont {{Phillips}}}, \bibinfo {author} {\bibfnamefont {R.~J.}\ \bibnamefont {{deBoer}}},\ and\ \bibinfo {author} {\bibfnamefont {S.~N.}\ \bibnamefont {{Paneru}}},\ }\href {https://doi.org/10.3389/fphy.2022.888476} {\bibfield  {journal} {\bibinfo  {journal} {Frontiers in Physics}\ }\textbf {\bibinfo {volume} {10}},\ \bibinfo {eid} {888476} (\bibinfo {year} {2022})},\ \Eprint {https://arxiv.org/abs/2112.12838} {arXiv:2112.12838 [nucl-th]} \BibitemShut {NoStop}%
\bibitem [{\citenamefont {{Foreman-Mackey}}\ \emph {et~al.}(2013)\citenamefont {{Foreman-Mackey}}, \citenamefont {{Hogg}}, \citenamefont {{Lang}},\ and\ \citenamefont {{Goodman}}}]{2013PASP..125..306F}%
  \BibitemOpen
  \bibfield  {author} {\bibinfo {author} {\bibfnamefont {D.}~\bibnamefont {{Foreman-Mackey}}}, \bibinfo {author} {\bibfnamefont {D.~W.}\ \bibnamefont {{Hogg}}}, \bibinfo {author} {\bibfnamefont {D.}~\bibnamefont {{Lang}}},\ and\ \bibinfo {author} {\bibfnamefont {J.}~\bibnamefont {{Goodman}}},\ }\href {https://doi.org/10.1086/670067} {\bibfield  {journal} {\bibinfo  {journal} {\pasp}\ }\textbf {\bibinfo {volume} {125}},\ \bibinfo {pages} {306} (\bibinfo {year} {2013})},\ \Eprint {https://arxiv.org/abs/1202.3665} {arXiv:1202.3665 [astro-ph.IM]} \BibitemShut {NoStop}%
\bibitem [{sup()}]{supplemental_material}%
  \BibitemOpen
  \href@noop {} {}\bibinfo {note} {{See Supplemental Material at [URL will be inserted by publisher] for additional details of the MCMC analysis.}}\BibitemShut {Stop}%
\bibitem [{\citenamefont {Schillebeeckx}\ \emph {et~al.}(2012)\citenamefont {Schillebeeckx}, \citenamefont {Becker}, \citenamefont {Danon}, \citenamefont {Guber}, \citenamefont {Harada}, \citenamefont {Heyse}, \citenamefont {Junghans}, \citenamefont {Kopecky}, \citenamefont {Massimi}, \citenamefont {Moxon}, \citenamefont {Otuka}, \citenamefont {Sirakov},\ and\ \citenamefont {Volev}}]{Schillebeeckx2012}%
  \BibitemOpen
  \bibfield  {author} {\bibinfo {author} {\bibfnamefont {P.}~\bibnamefont {Schillebeeckx}}, \bibinfo {author} {\bibfnamefont {B.}~\bibnamefont {Becker}}, \bibinfo {author} {\bibfnamefont {Y.}~\bibnamefont {Danon}}, \bibinfo {author} {\bibfnamefont {K.}~\bibnamefont {Guber}}, \bibinfo {author} {\bibfnamefont {H.}~\bibnamefont {Harada}}, \bibinfo {author} {\bibfnamefont {J.}~\bibnamefont {Heyse}}, \bibinfo {author} {\bibfnamefont {A.}~\bibnamefont {Junghans}}, \bibinfo {author} {\bibfnamefont {S.}~\bibnamefont {Kopecky}}, \bibinfo {author} {\bibfnamefont {C.}~\bibnamefont {Massimi}}, \bibinfo {author} {\bibfnamefont {M.}~\bibnamefont {Moxon}}, \bibinfo {author} {\bibfnamefont {N.}~\bibnamefont {Otuka}}, \bibinfo {author} {\bibfnamefont {I.}~\bibnamefont {Sirakov}},\ and\ \bibinfo {author} {\bibfnamefont {K.}~\bibnamefont {Volev}},\ }\href {https://doi.org/10.1016/j.nds.2012.11.005} {\bibfield  {journal} {\bibinfo  {journal} {Nuclear Data Sheets}\ }\textbf {\bibinfo {volume} {113}},\ \bibinfo {pages} {3054}
  (\bibinfo {year} {2012})}\BibitemShut {NoStop}%
\bibitem [{\citenamefont {Larson}\ \emph {et~al.}(1983)\citenamefont {Larson}, \citenamefont {Larson}, \citenamefont {Harvey}, \citenamefont {Hill},\ and\ \citenamefont {Johnson}}]{ENDF333}%
  \BibitemOpen
  \bibfield  {author} {\bibinfo {author} {\bibfnamefont {D.}~\bibnamefont {Larson}}, \bibinfo {author} {\bibfnamefont {N.}~\bibnamefont {Larson}}, \bibinfo {author} {\bibfnamefont {J.}~\bibnamefont {Harvey}}, \bibinfo {author} {\bibfnamefont {N.}~\bibnamefont {Hill}},\ and\ \bibinfo {author} {\bibfnamefont {C.}~\bibnamefont {Johnson}},\ }\href {https://www.nndc.bnl.gov/endfdocs/} {\emph {\bibinfo {title} {ENDF-333, Application of New Techniques to ORELA Neutron Transmission Measurementsand their Uncertainty Analysis: The Case of Natural Nickel from 2 keV to 20 MeV}}},\ \bibinfo {type} {Tech. Rep.}\ \bibinfo {number} {ORNL/TM-8203}\ (\bibinfo  {institution} {Oak Ridge National Laboratory},\ \bibinfo {year} {1983})\BibitemShut {NoStop}%
\bibitem [{\citenamefont {Junghans}\ \emph {et~al.}(2025)\citenamefont {Junghans}, \citenamefont {Beyer}, \citenamefont {Capote}, \citenamefont {Bemmerer}, \citenamefont {Hensel}, \citenamefont {Kögler}, \citenamefont {Müller}, \citenamefont {Römer}, \citenamefont {Schmidt}, \citenamefont {Schwengner}, \citenamefont {Turkat}, \citenamefont {Turko},\ and\ \citenamefont {Urlaß}}]{Junghans2025}%
  \BibitemOpen
  \bibfield  {author} {\bibinfo {author} {\bibfnamefont {A.}~\bibnamefont {Junghans}}, \bibinfo {author} {\bibfnamefont {R.}~\bibnamefont {Beyer}}, \bibinfo {author} {\bibfnamefont {R.}~\bibnamefont {Capote}}, \bibinfo {author} {\bibfnamefont {D.}~\bibnamefont {Bemmerer}}, \bibinfo {author} {\bibfnamefont {T.}~\bibnamefont {Hensel}}, \bibinfo {author} {\bibfnamefont {T.}~\bibnamefont {Kögler}}, \bibinfo {author} {\bibfnamefont {S.}~\bibnamefont {Müller}}, \bibinfo {author} {\bibfnamefont {K.}~\bibnamefont {Römer}}, \bibinfo {author} {\bibfnamefont {K.}~\bibnamefont {Schmidt}}, \bibinfo {author} {\bibfnamefont {R.}~\bibnamefont {Schwengner}}, \bibinfo {author} {\bibfnamefont {S.}~\bibnamefont {Turkat}}, \bibinfo {author} {\bibfnamefont {J.}~\bibnamefont {Turko}},\ and\ \bibinfo {author} {\bibfnamefont {S.}~\bibnamefont {Urlaß}},\ }\href {https://doi.org/https://doi.org/10.1016/j.anucene.2025.111363} {\bibfield  {journal} {\bibinfo  {journal} {Annals of Nuclear Energy}\ }\textbf {\bibinfo {volume} {218}},\
  \bibinfo {pages} {111363} (\bibinfo {year} {2025})}\BibitemShut {NoStop}%
\end{thebibliography}%

\end{document}